\documentclass[aps,prb,superscriptaddress,preprintnumbers,
showpacs,legalpaper,twoside,twocolumn,amsmath,amssymb]{revtex4}
\usepackage{graphicx}
\usepackage{psfrag}
\usepackage{color}
\usepackage{subfigure}
\usepackage{amsmath}
\definecolor{dred}{rgb}{0.7,0.0,0.0}

\begin{document}

\title{ Magnetic and Metallic State at Intermediate Hubbard $U$ Coupling \\
in Multiorbital Models for Undoped Fe Pnictides}

\author{Rong Yu}
\affiliation{Department of Physics and Astronomy, The University of Tennessee, Knoxville, TN 37996}
\affiliation{Materials Science and Technology Division, Oak Ridge National Laboratory, Oak Ridge, TN 32831}

\author{Kien T. Trinh}
\affiliation{Department of Physics and Astronomy, University of Southern California, Los Angeles, CA 90089}

\author{Adriana Moreo}
\affiliation{Department of Physics and Astronomy, The University of Tennessee, Knoxville, TN 37996}
\affiliation{Materials Science and Technology Division, Oak Ridge National Laboratory, Oak Ridge, TN 32831}

\author{Maria Daghofer}
\affiliation{Department of Physics and Astronomy, The University of Tennessee, Knoxville, TN 37996}
\affiliation{Materials Science and Technology Division, Oak Ridge National Laboratory, Oak Ridge, TN 32831}


\author{Jos\'e A. Riera}
\affiliation{Instituto de F\'{\i}sica Rosario, Consejo Nacional de
Investigaciones Cient\'{\i}ficas y T\'ecnicas,
Universidad Nacional de Rosario, 2000 Rosario, Argentina}

\author{Stephan Haas}
\affiliation{Department of Physics and Astronomy, University of Southern California, Los Angeles, CA 90089}

\author{Elbio Dagotto}
\affiliation{Department of Physics and Astronomy, The University of Tennessee, Knoxville, TN 37996}
\affiliation{Materials Science and Technology Division, Oak Ridge National Laboratory, Oak Ridge, TN 32831}

\date{\today}

\begin{abstract}
Multi-orbital Hubbard model Hamiltonians for the undoped parent compounds of the
Fe-pnictide superconductors are here investigated using mean-field techniques. For a realistic four-orbital model,
our results show the existence of an intermediate Hubbard $U$ coupling regime where the
mean-field ground state has spin stripe magnetic order, as in neutron scattering experiments,
while remaining metallic, due to the phenomenon of  band overlaps. The angle-resolved
photoemission intensity and Fermi surface of this magnetic and metallic state are discussed.
Other models are also investigated, including a two orbital model where not only the mean-field technique
can be used, but also Exact Diagonalization in small clusters and the Variational Cluster Approximation in the
bulk. The combined results of the three techniques point toward the existence of an intermediate-coupling magnetic and
metallic state in the two-orbital model, similar to the intermediate coupling
mean-field state of the four-orbital model. We conclude that the state discussed here is compatible with
the experimentally known properties of the undoped Fe-pnictides.
\end{abstract}


\maketitle

\section{Introduction}

The discovery of superconductivity in the Fe pnictides has opened an area of
research that is attracting considerable attention.\cite{Fe-SC, chen1, chen2, wen, chen3, 55, ren1, ren2}
In the early stages of these investigations, the layered structure of the Fe pnictides
superconductors,\cite{Fe-SC, chen1, chen2, wen, chen3, 55, ren1, ren2}
the existence of a magnetic spin striped state revealed by neutron scattering in the undoped limit,\cite{neutrons1,neutrons2}
and their large superconducting critical temperatures~\cite{Fe-SC, chen1, chen2, wen, chen3, 55, ren1, ren2}
motivated discussions on a possible close relation between these new
Fe-based materials and the high-temperature cuprate superconductors.
However, it was clear from the initial investigations that there were substantial differences as well: for example,
the resistivity vs. temperature curves of the parent compounds~\cite{Fe-SC, chen1, chen2, wen, chen3, 55, ren1, ren2} do not show the characteristic
Mott gapped behavior of, e.g., LaCuO$_4$. In fact LaOFeAs behaves
as a bad metal or semiconductor,\cite{Fe-SC, chen1, chen2, wen, chen3, 55, ren1, ren2} but not as an insulator. Moreover, the magnetic moment in the spin striped
state of LaOFeAs is
much smaller than expected.\cite{neutrons1,neutrons2}
Although further neutron scattering research has shown that the magnetic order parameters
are larger in other Fe pnictides,\cite{neutrons4}
their values are still below those anticipated from
band structure calculations~\cite{singh,first, xu, cao, fang2}
or from the large Hubbard $U$ limit of model Hamiltonians~\cite{daghofer} (unless couplings
are in a  spin frustrated regime~\cite{si}). In summary, the parent compounds of the
Fe superconductors behave in a manner different
from the parent compounds of the Cu-oxide superconductors
because the zero temperature resistivity is finite and the magnetic order weak. However, the pnictides
are also different
from BCS materials, where the normal state is a non-magnetic metal with low resistivity.
Then, the Fe superconductors appear to be
in an {\it intermediate} regime of couplings, somewhere in between, e.g., MgB$_2$ and the Cu oxide
superconductors.\cite{Jaro,basov} The ``antiferromagnetic metallic''
nature of LaOFeAs is clearly different from an
antiferromagnetic insulator or a non-magnetic metal.
Further confirming the need to focus on the intermediate coupling regime for the Fe pnictides parent compounds,
a pseudogap has been observed~\cite{Y.Ishida,T.Sato,liu2,L.Zhao} in their density of states (DOS), which is
different from the featureless DOS of
a good metal or the gapped DOS of an insulator.

Much of the current theoretical
effort~\cite{daghofer,si,kuroki,mazin,FCZhang,han,korshunov,baskaran,yao,
  xu2,plee,yildirim,scalapino,hu,zhou,lorenzana,sk,parish,choi,wang,yang,shi,2orbitals,calderon}
has focused thus far on two well defined limits. On one hand,
band structure calculations have reported the existence of Fermi pockets,\cite{singh,first, xu, cao, fang2}
which were confirmed by
photoemission
experiments.\cite{hashimoto,arpes,arpes2,C.Martin,T.Chen,parker,arpes3}
On the other hand,
intuition on the physics of the model Hamiltonians is often gained
by investigating the large $U$ regime.
Although results by some of us for a two-orbital model
using numerical techniques
have already shown that a small magnetic order can be accommodated
at intermediate couplings,\cite{daghofer}
this range of couplings
is typically the most difficult to handle using computer simulations
and, moreover, more bands are expected to be of relevance for a better quantitative
description of the Fe pnictides.

In this publication, multi-band Hubbard models are investigated
using mean-field and numerical techniques. Our most important result is the discovery of an
intermediate Hubbard $U$ coupling regime where the ground state is an antiferromagnetic metal.
More specifically, for $U$ larger than a critical value $U_{\rm c1}$
the spin striped order develops with continuity from zero.
In spite of a gap at particular momenta, the overall state remains metallic (there is a nonzero
weight at the chemical potential in the DOS) due
to the phenomenon of band overlaps. Further increasing the coupling to $U_{\rm c2}$, a fully gapped
insulator is stabilized. Thus, the intermediate regime $U_{\rm c1}$$<$$U$$<$$U_{\rm c2}$ is
simultaneously (i) magnetic with a small order parameter and (ii) metallic. Moreover,
the DOS reveals the existence of a pseudogap in this regime. All these properties are compatible with our
current knowledge of the Fe pnictides parent compounds.

While these results are interesting, they were obtained using mean-field approximations. For the
realistic case of four or five orbitals, it is difficult to obtain reliable
numerical results
to confirm the mean-field predictions.
However, for the case of two orbitals, calculations can be carried out using
simultaneously the mean-field technique, that also reveals
an intermediate coupling regime similar to that
of the four and five orbitals models,
together with the Exact Diagonalization (ED)\cite{RMP} and Variational Cluster Approximation (VCA)\cite{Aic03,Pot03} methods used
before.\cite{daghofer} Below, it is reported that the results using these
computational methods are compatible with
those of the mean-field approximation, providing
confidence that the mean-field method
may have captured the essence of the problem. However, it is certainly desirable that
future investigations confirm our main results.

For completeness, here some related previous literature is briefly mentioned.
The band overlap mechanism for an insulator to metal transition has been extensively studied before using band structure
calculations in a variety of contexts,
such as TlCl and TlBr,\cite{samara} solid hydrogen,\cite{hydrogen}
and bromine under high pressure.\cite{bromine}
%
Closer to the present results, the existence
of an intermediate Hubbard $U$ regime with an antiferromagnetic metallic
state was previously discussed by Duffy and one of the authors (A.M.)~\cite{duffy}
in the context of the high temperature Cu-oxide superconductors and using the one-band Hubbard
model, after introducing hopping terms $t'$ and $t''$ between next-nearest-neighbor sites. Density functional methods
were also used before to discuss AF-metallic states and band-overlap insulator-metal transitions.\cite{sander,callaway}
Within dynamical mean-field theory, an AF-metallic state has also been discussed.\cite{marcelo}
Experimentally, itinerant antiferromagnetic states were found in the pyrite NiS$_{2-x}$Se$_x$,\cite{miyasaka,niklowitz}
in heavily doped manganites,\cite{moritomo} in ruthenates,\cite{cao97}
in organic conductors,\cite{organic,organic2} and in several other materials.
An incommensurate spin density wave was also reported in
metallic V$_{2-y}$O$_3$.\cite{bao}
The results discussed in this manuscript establish an interesting connection
between the Fe pnictides and the materials mentioned in this paragraph.

The organization of the paper is as follows. In Section II,
results for a four-orbital model are presented. This
includes a discussion of the model, the mean-field technique, and the
results, with emphasis on the intermediate coupling
state. The photoemission predictions for this state are discussed. In
Section III, we show results for the  two-orbital model using
ED\cite{RMP} and  VCA methods\cite{Aic03,Pot03} in addition
to mean-field approximations.
Section IV contains our main conclusions.

\section{Results for a four-orbital model}

In this section, we will describe a possible minimal four-orbital model for
the Fe-based superconductors. This model presents a Fermi surface similar to that
obtained with band structure calculations.

\subsection{The four-orbital model}

Previous studies have suggested that the Fe-As planes are the most
important substructures of the full crystal that must be analyzed in order
to reproduce the physical properties of the Fe pnictides
close to the Fermi surface. We consequently focus on these
Fe-As planes in the present study.
The effective Fe-Fe hopping Hamiltonian, using As as a bridge,
can be obtained within the
framework of the Slater-Koster (SK) formalism.\cite{slater} By this
procedure, here we will construct a minimal model defined on a Fe square lattice,
consisting of the four Fe $d$ orbitals $xz$, $yz$, $xy$, and
$x^2-y^2$. It is assumed that the $d_{3z^2-r^2}$ orbital lies at a
substantially lower energy and is thus
always filled with two electrons.\cite{singh,first, xu, cao, fang2} While the SK procedure
is not as quantitatively accurate as a full band-structure calculation, it can
still provide the proper model Hamiltonian, because it correctly takes
into account the geometry of the system and illustrates which orbitals are
connected to one another at different lattice sites. Thus, our procedure here will be to use the SK method
to construct the formal model, and then obtain the actual numerical values of the
hopping parameters via comparison with band-structure calculations.

The Hamiltonian $H=H_0 + H_{\rm int}$ includes two parts: the hopping term $H_0$ and the
interaction term $H_{\rm int}$. The hopping term in real space reads
\begin{eqnarray}\label{E.H0r}
H_0 &=& \sum_{\bf i,j}\sum_{\mu,\nu}\sum_\sigma \left( T^{\mu,\nu}_{\bf i,j}
d^\dagger_{{\bf i},\mu,\sigma} d_{{\bf j},\nu,\sigma} + h.c. \right),
\end{eqnarray}
where $d^\dagger_{{\bf i},\mu,\sigma}$ creates an electron
at site ${\bf i}$
with spin $\sigma$ on the $\mu$-th orbital ($\mu=1,2,3,4$ stands for the
$xz$, $yz$, $xy$, and $x^2$-$y^2$ orbitals, respectively). Here, hoppings
at nearest-neighbors (NN) and also at next-nearest-neighbors (NNN) along the plaquette
diagonals were considered. The hopping
tensor $T^{\mu,\nu}_{\bf i,j}$ has a complicated real-space structure
that will not be reproduced here. $H_0$ has a
simpler form when transformed to momentum space:

\begin{eqnarray}\label{E.H0k}
H_0 &=& \sum_{\mathbf{k}} \sum_{\mu,\nu}\sum_\sigma T^{\mu,\nu}(\mathbf{k})
d^\dagger_{\mathbf{k},\mu,\sigma} d_{\mathbf{k},\nu,\sigma},
\end{eqnarray}
with
\begin{eqnarray}
T^{11} &=& -2t_2\cos k'_x -2t_1\cos k'_y -4t_3 \cos k'_x \cos k'_y, \label{eq:t11}\\
T^{22} &=& -2t_1\cos k'_x -2t_2\cos k'_y -4t_3 \cos k'_x \cos k'_y, \label{eq:t22}\\
T^{12} &=& -4t_4\sin k'_x \sin k'_y, \label{eq:t12}\\
T^{33} &=& -2t_5(\cos(k'_x+\pi)+\cos(k'_y+\pi)) \nonumber\\
       & & -4t_6\cos(k'_x+\pi)\cos(k'_y+\pi) +\Delta_{xy}, \\
T^{13} &=& -4it_7\sin k'_x + 8it_8\sin k'_x \cos k'_y, \\
T^{23} &=& -4it_7\sin k'_y + 8it_8\sin k'_y \cos k'_x, \\
T^{44} &=& -2t_{17}(\cos(k'_x+\pi)+\cos(k'_y+\pi)) \nonumber\\
       & & -4t_9\cos(k'_x+\pi)\cos(k'_y+\pi) +\Delta_{x^2-y^2}, \\
T^{14} &=& -4it_{10}\sin k'_y, \\
T^{24} &=& ~~4it_{10}\sin k'_x, \\
T^{34} &=& ~~0.
\end{eqnarray}

\begin{figure}[h]
\begin{center}
\vskip -0.5cm
\centerline{\includegraphics[width=8.0cm,clip,angle=0]{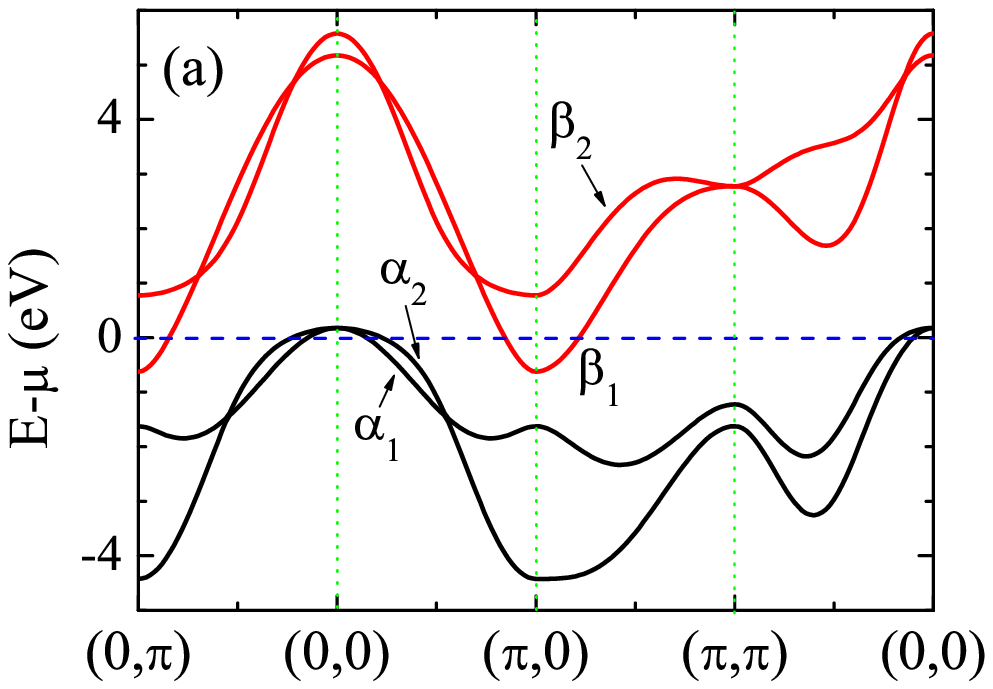}}
\vskip -0.7cm
\centerline{\includegraphics[width=6.5cm,clip,angle=0]{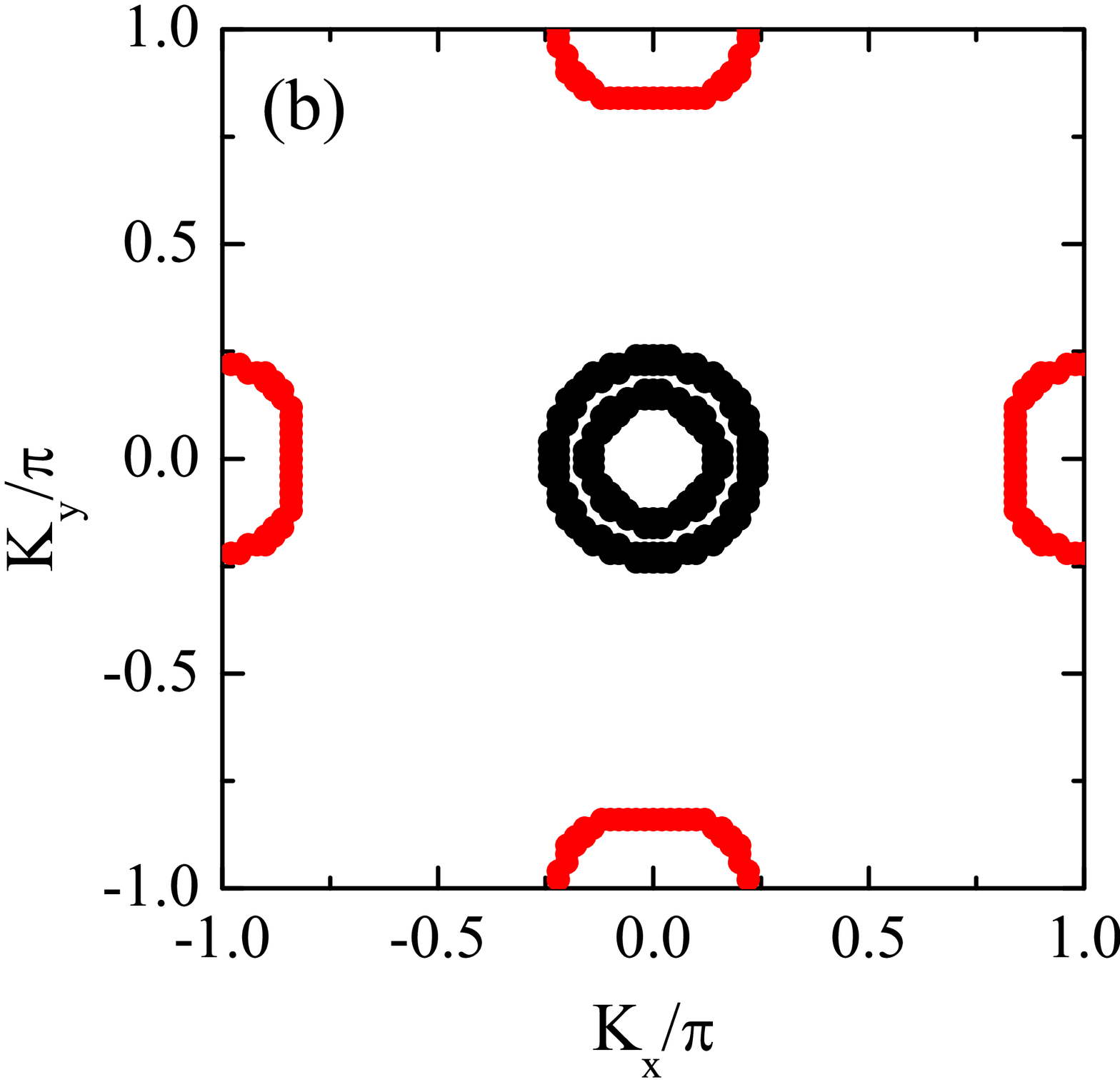}}
\vskip -0.7cm
\caption{(Color online) (a) Band structure corresponding to the four-orbital
tight-binding Hamiltonian Eq.~(\ref{E.H0k}), using the LDA fitted values of the
hopping parameters provided in Table~\ref{T.hoppara}. The Fermi surface
is formed by two hole-like bands ($\alpha_1$, $\alpha_2$) and
one electron-like band ($\beta_1$). The chemical potential is at
 $0$. (b) The topology of the corresponding Fermi surface. }
\label{F.BandU0.0}
\vskip -1.0cm
\end{center}
\end{figure}

\begin{figure}[h]
\begin{center}
\vskip -0.5cm
\centerline{\includegraphics[width=8.0cm,clip,angle=0]{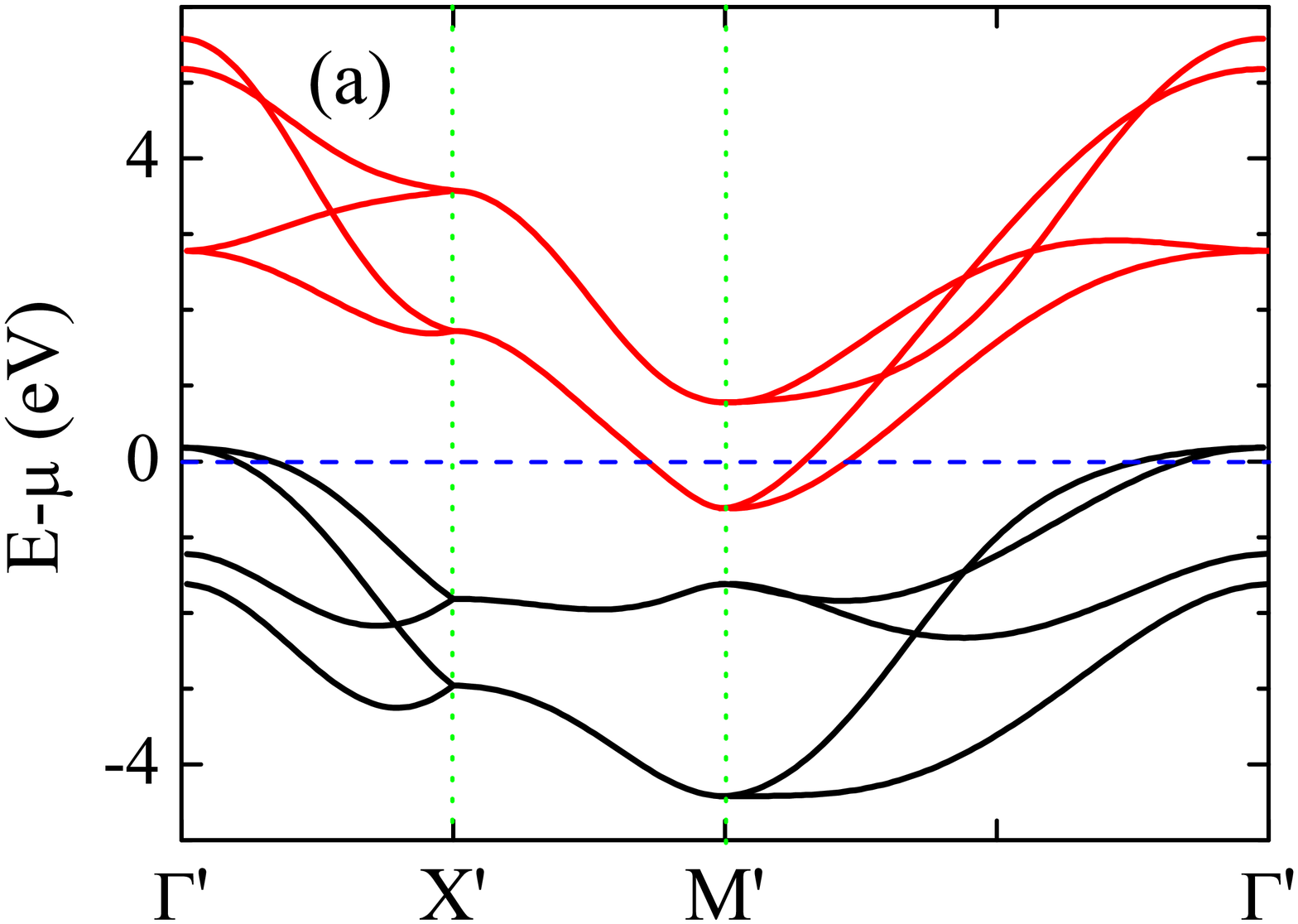}}
\vskip -1.0cm
\centerline{\includegraphics[width=6.5cm,clip,angle=0]{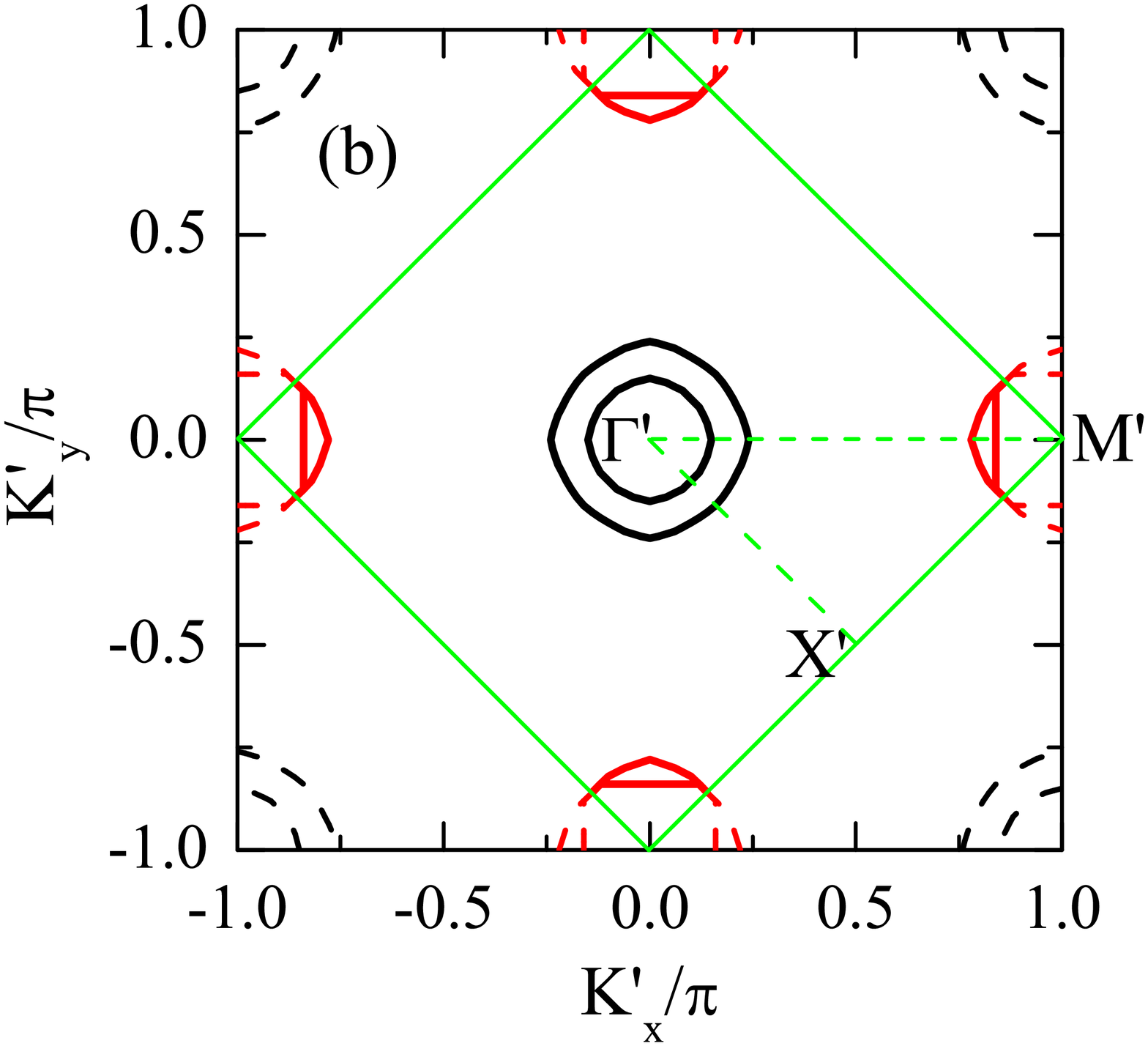}}
\vskip -0.7cm
\caption{(Color online)
(a) Band structure of the eight-orbital problem in the reduced BZ
obtained by ``folding'' the results presented in Fig.~\ref{F.BandU0.0}(a).
(b) Fermi surface
of the eight-orbital problem obtained by folding the FS obtained in
Fig.~\ref{F.BandU0.0}(b). The FS in
the first (second) BZ are indicated by continuous (dashed) lines.
}
\label{F.Band-bis}
\vskip -1.0cm
\end{center}
\end{figure}

Equation~(\ref{E.H0k}) represents a
matrix in the basis $\{d^{\dagger}_{{\bf k},\mu,\sigma}\}$ where
${\bf k=k'}$ if $\mu=1$ or 2 and ${\bf k=k'+Q}$ if $\mu=3$ or 4 with
$-\pi<k_x, k_y\leq\pi$, ${\bf k'}$ is defined
in the reduced
Brillouin zone corresponding to the two-Fe unit cell, and
$\mathbf{Q}=(\pi,\pi)$.
In other words, the above expressions
couple states with momentum
$\mathbf{k'}$ for orbitals $xz$ and $yz$ to states with
momentum $\mathbf{k'+Q}$ for orbitals $xy$ and $x^2-y^2$.
The momentum $\mathbf{Q}$ appears
after considering the staggered location
of the As atoms above and below the plane defined by the Fe atoms.
As already mentioned,
the mathematical form of this model arises directly from
the Slater-Koster considerations.
This problem is equivalent to an eight-orbital model with
a Hamiltonian expanded in the basis $\{d^{\dagger}_{{\bf k'},\mu,\sigma},
d^{\dagger}_{{\bf k'+Q},\mu,\sigma}\}$.
As a result, the eight-orbital
band structure and Fermi surface (Fig.~\ref{F.Band-bis})
are obtained by ``folding'' the results in
the equivalent four-orbital problem (Fig.~\ref{F.BandU0.0}).

The actual values of the hopping parameters could in principle be
obtained from the overlap integrals in the SK
formalism.\cite{slater,Moreoetal08}
However, to properly reproduce the Fermi surface obtained in the
Local Density Approximation (LDA)~\cite{singh,first, xu, cao, fang2} it is better to fit
the values of those hoppings. The
parameters used, as well as the on-site energies $\Delta_\mu$ for the
$xy$ and $x^2-y^2$ orbitals, are
listed in Table~\ref{T.hoppara}. The on-site energy term is
given by $\sum_{{\bf i},\mu} \Delta_{\mu} n^{\mu}_{\bf i}$ (standard notation)
and it is part of the tight-binding Hamiltonian.

\begin{table}[h]
\caption{Fitted hopping parameters and on-site energies for the
four-orbital model used in this section (in eV units).}\label{T.hoppara} \centering
\vskip 0.3cm
\begin{tabular}{||c|c||c|c||}
\hline
$\Delta_{xy}$ & -0.600 & $\Delta_{x^2-y^2}$ & -2.000 \\
\hline
$t_1$ & ~0.500 & $t_2$ & ~0.150 \\
\hline
$t_3$ & -0.175 & $t_4$ & -0.200 \\
\hline
$t_5$ & ~0.800 & $t_6$ & -0.450 \\
\hline
$t_7$ & ~0.460 & $t_8$ & ~0.005 \\
\hline
$t_9$ & -0.800 & $t_{10}$ & -0.400 \\
\hline
$t_{17}$ & ~0.900 & & \\
\hline
\end{tabular}
\end{table}


In Figs.~\ref{F.BandU0.0} and
\ref{F.Band-bis}, we show the band structure and the
corresponding Fermi surface for the four-orbital tight-binding
Hamiltonian in Eq.~(\ref{E.H0k}), using Table~\ref{T.hoppara}. Since we assume that the
$3z^2-r^2$ orbital is always doubly occupied, the chemical potential
in the undoped case is determined by locating $n=4$ electrons per site in the
four bands considered here. As shown in Fig.~\ref{F.BandU0.0}(b),
two hole pockets centered
at $(0,0)$ (arising from the $\alpha_1$ and $\alpha_2$ bands) and
four pieces of
two electron pockets centered at $(0,\pi)$ and $(\pi,0)$ (from the
$\beta_1$ band) are obtained. The shape of the Fermi surface
qualitatively reproduces the band-structure
LDA calculations~\cite{singh,first, xu, cao, fang2}
after a $45^o$ rotation about the center of the First Brillouin Zone
(FBZ), as presented in Fig.~\ref{F.Band-bis}(b), due to the rotation
from the Fe-Fe axis to the Fe-As axis.

\begin{figure}[h]
\vskip -0.3cm
\centerline{\includegraphics[width=9cm,clip,angle=0]{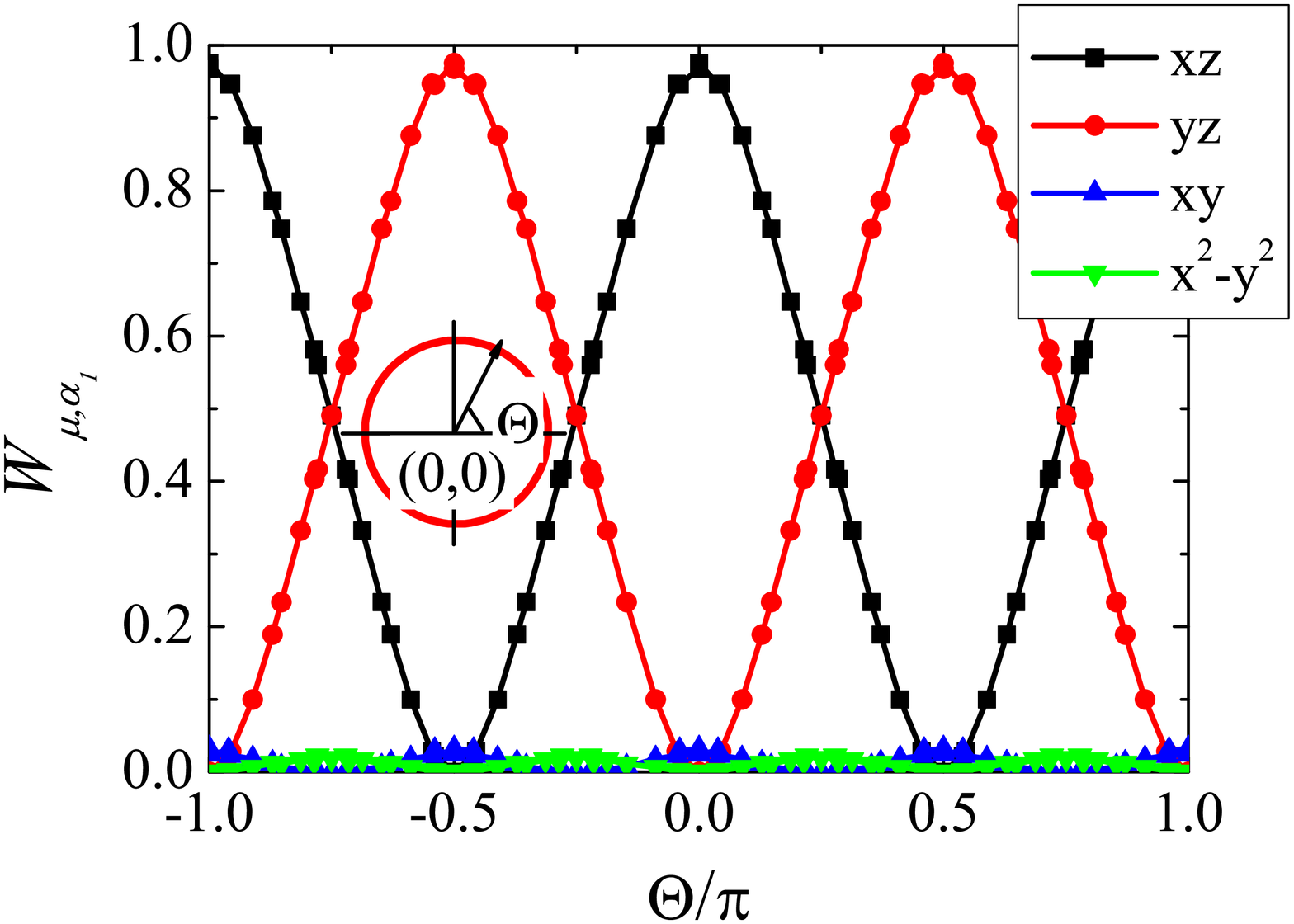}}
\vskip -0.5cm
%
\caption{(Color online) The projected orbital weight
$W_{\mu,\lambda}$ of states at the Fermi surface. Shown, as example, are results for the
outer hole pocket
centered at $(0,0)$.
The
definition of $\Theta$ is given in the inset.
}
\vskip -0.3cm
\label{F.Project}
\end{figure}

To study the relation between the orbital hybridization and the Fermi surface topology,
the projected weight of each orbital at both the hole and
electron pockets were calculated. These weights are defined via the eigenvectors
of $H_0$: $W_{\mu,\lambda} (\mathbf{k}) =
\frac{1}{2} \sum_\sigma |U_{\mathbf{k},\mu,\sigma;\lambda}|^2$,
where $\lambda$ denotes the band index ($\alpha_1, \alpha_2,
\beta_1, \beta_2$), and $\mu$ refers to the four $d$ orbitals. The matrix
$U_{\mathbf{k},\mu,\sigma;\lambda}$ diagonalizes the system (see Eq.~(\ref{diago}) below).
An example of the angle-resolved weights in momentum space are shown in
Fig.~\ref{F.Project}. The two hole pockets centered at
$(0,0)$ mostly arise from the $xz$ and $yz$ orbitals, compatible with LDA~\cite{singh,first, xu, cao, fang2} and with
much simpler descriptions
based only on two orbitals.\cite{scalapino,daghofer} The electron
pocket centered at $(\pi,0)$ ($(0,\pi)$) arises mainly from the
hybridization of the $xz$ ($yz$) and $xy$ orbitals (not shown). These results are also
qualitatively consistent with those from the first-principles
calculations.\cite{fang2} However, there are some quantitative discrepancies that lead us
to believe that probably longer-range than NNN
plaquette-diagonal hoppings are needed to fully reproduce the
LDA results including orbital weights. Nevertheless, the discussion below on the
metallic magnetic phase at intermediate couplings is robust,
and we believe it will survive when more complex
multi-orbital models are used in the future.

Note that the eigenenergies (band dispersion) along
the $(0,0)\rightarrow(\pi,0)$ and
$(0,0)\rightarrow(0,\pi)$ directions are symmetric about $(0,0)$, but the
eigenvectors ($W_{\mu,\lambda}$) show a large anisotropy. For
instance, at the Fermi level the $\alpha_1$ band is almost $xz$-like
along the $(0,0)\rightarrow(\pi,0)$ direction but almost $yz$-like
along the $(0,0)\rightarrow(0,\pi)$ direction. Below, it will be discussed how
this anisotropy affects the mean-field results for the
interacting system.

Let us now consider the interaction term,\cite{daghofer} which reads
\begin{eqnarray}\label{E.Hint}
H_{\rm int} &=& U\sum_{{\bf i},\mu}n_{{\bf i},\mu,\uparrow}n_{{\bf i},\mu,\downarrow}
+(U'-{J\over{2}})\sum_{{\bf i},\mu\neq\nu} n_{{\bf i},\mu}n_{{\bf i},\nu}
\nonumber \\
& & -2J\sum_{{\bf i},\mu\neq\nu}\mathbf{S}_{{\bf i},\mu}\cdot\mathbf{S}_{{\bf i},\nu},
\end{eqnarray}
where $\mathbf{S}_{{\bf i},\mu}$ ($n_{{\bf i},\mu}$) is the spin (charge
density) of orbital $\mu$ at site ${\bf i}$, and
$n_{{\bf i},\mu}=n_{{\bf i},\mu,\uparrow}+n_{{\bf i},\mu,\downarrow}$. The first term
is a Hubbard repulsion for the electrons in the same orbital. The
second term describes an on-site inter-orbital repulsion, where the
standard relation $U'=U-J/2$ caused by rotational invariance is used.\cite{RMP01} The last term in
Eq.~(\ref{E.Hint}) is a Hund term with a ferromagnetic coupling $J$.
A complete description would also require a
pair-hopping interaction similar to the last term of
Eq.~(\ref{eq:Hint2}), where the interaction term for the
two-orbital model is shown. But
ED was used to test its impact in the case of two orbitals,
and it was not found to be important. Consequently, it was neglected in the
mean field treatment.

\subsection{The mean-field approach}
To study the ground state properties of the system, we
apply a mean-field approximation to the model Hamiltonian described
by Eqs.~(\ref{E.H0r}) to (\ref{E.Hint}).
We follow here the simple standard assumption of considering only the
mean-field values for the diagonal operators:\cite{nomura}

\begin{eqnarray}\label{E.MFA}
\langle d^\dagger_{{\bf i},\mu,\sigma} d_{{\bf j},\nu,\sigma'}\rangle =
\left(n_\mu+\frac{\sigma}{2}\cos(\mathbf{q}\cdot\mathbf{r}_{\bf i})m_\mu\right)
\delta_{\bf ij}\delta_{\mu\nu}\delta_{\sigma\sigma'},
\end{eqnarray}
where $\mathbf{q}$ is the ordering vector of the possible magnetic
order. $n_\mu$ and $m_\mu$ are mean-field parameters describing the
charge density and magnetization of the orbital $\mu$, and the rest of the
notation is standard. Applying
Eq.~(\ref{E.MFA}) to $H_{\rm int}$, the mean-field Hamiltonian in
momentum space can be written as

\begin{eqnarray}\label{E.HMF}
H_{\rm MF} = H_0 + C + \sum_{\mathbf{k},\mu,\sigma}
\epsilon_\mu d^\dagger_{\mathbf{k},\mu,\sigma}
d_{\mathbf{k},\mu,\sigma}\nonumber\\
+ \sum_{\mathbf{k},\mu,\sigma} \eta_{\mu,\sigma}
 (d^\dagger_{\mathbf{k},\mu,\sigma} d_{\mathbf{k+q},\mu,\sigma} +
d^\dagger_{\mathbf{k+q},\mu,\sigma} d_{\mathbf{k},\mu,\sigma}),
\end{eqnarray}
where $\mathbf{k}$ runs over the extended FBZ, $H_0$ is
the hopping term in Eq.~(\ref{E.H0k}),
\begin{eqnarray}
C=&-&NU\sum_{\mu}\left(n^2_\mu-\frac{1}{4}m^2_\mu\right)
- N(2U'-J)\sum_{\mu\neq\nu}n_\mu n_\nu \nonumber \\
&+& \frac{NJ}{4} \sum_{\mu\neq\nu} m_\mu m_\nu \nonumber
\end{eqnarray}
is a constant, $N$ the lattice size, and we used the definitions
\begin{eqnarray}
\epsilon_\mu = Un_\mu + (2U'-J)\sum_{\nu\neq\mu}
n_\nu, \\
\eta_{\mu,\sigma} =
-\frac{\sigma}{2}\left(Um_\mu+J\sum_{\nu\neq\mu}m_\nu\right).
\end{eqnarray}

The above mean-field Hamiltonian can be numerically solved for a fixed
set of mean-field parameters using standard
library subroutines.
The parameters $n_\mu$ and $m_\mu$ are obtained in a
self-consistent manner by minimizing the energy. In practice an
initial guess for $n_\mu$ and $m_\mu$ serves as a set of
input parameters for a given value of the
couplings $U$ and $J$.
The mean-field Hamiltonian is then diagonalized
and $n_\mu$ and $m_\mu$ are reevaluated using Eq.~(\ref{E.MFA}).
This procedure is iterated until both $n_\mu$ and $m_\mu$ have converged.
During the iterative procedure
$\sum_\mu n_\mu$=$4$ was enforced at each step, such that
the total charge density is a constant. Note
that in the mean-field approximation an electron with momentum
$\mathbf{k}$ is coupled to an electron with momentum $\mathbf{k+q}$,
where $\mathbf{q}$ is the vector associated with the magnetic ordering.
Then, the Hamiltonian in Eq.~(\ref{E.HMF}) is solved
only in the magnetic reduced Brillouin
Zone with only half of the size of the unfolded FBZ.

The numerical solution of the mean-field Hamiltonian immediately allows for the
evaluation of the band structure, the density of states (DOS), and the
magnetization ($M=\sum_\mu m_\mu$) at the ordering wavevector
$\mathbf{q}$. Moreover, we can also calculate the photoemission
spectral function. Assuming that the mean-field Hamiltonian
Eq.~(\ref{E.HMF}) is diagonalized by the unitary transformation
\begin{eqnarray}\label{diago}
d_{\mathbf{k},\mu,\sigma} = \sum_\lambda
U_{\mathbf{k},\mu,\sigma;\lambda} \gamma_\lambda, \\
H_{\rm MF} = \sum_{\lambda}
\rho_\lambda\gamma^\dagger_\lambda\gamma_\lambda,
\end{eqnarray}
then the spectral function is given by
\begin{eqnarray}\label{E.Akw}
A(\mathbf{k},\omega) = \sum_\lambda\sum_{\mu,\nu,\sigma}
U_{\mathbf{k},\mu,\sigma;\lambda}
U^\dagger_{\lambda;\mathbf{k},\nu,\sigma}
\delta(\omega-\rho_\lambda).
\end{eqnarray}
In practice $\delta(\omega-\rho_\lambda)$ was here substituted by
$\frac{1}{\pi}\frac{\varepsilon}{\varepsilon^2+(\omega-\rho_\lambda)^2}$
with a broadening $\varepsilon=0.05$~eV.

\subsection{Mean-field results}
In this subsection, the mean-field results for the
four-orbital model previously described are presented. Experimentally,
a (bad) metallic phase with spin-stripe magnetic order at
wavevectors $(0,\pi)$ was observed in the undoped compound
LaOFeAs.\cite{neutrons1,neutrons2} It is then important to investigate
in the mean-field approximation the properties of such a
spin striped ordered state. Here
we show the numerical results for the mean-field approximation defined
on a $100\times100$ square lattice.

\subsubsection{Magnetic order}\label{sec:megn_order4}

\begin{figure}[h]
\begin{center}
\vskip -0.6cm
\centerline{\includegraphics[width=9cm,clip,angle=0]{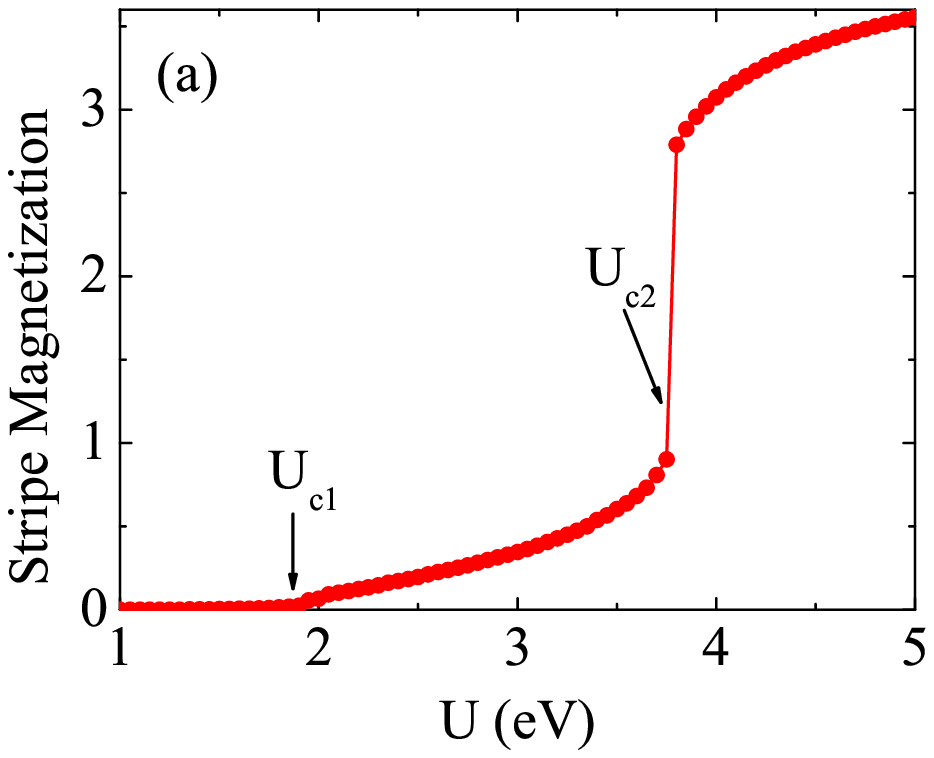}}
\vskip -0.7cm
\centerline{\includegraphics[width=9cm,clip,angle=0]{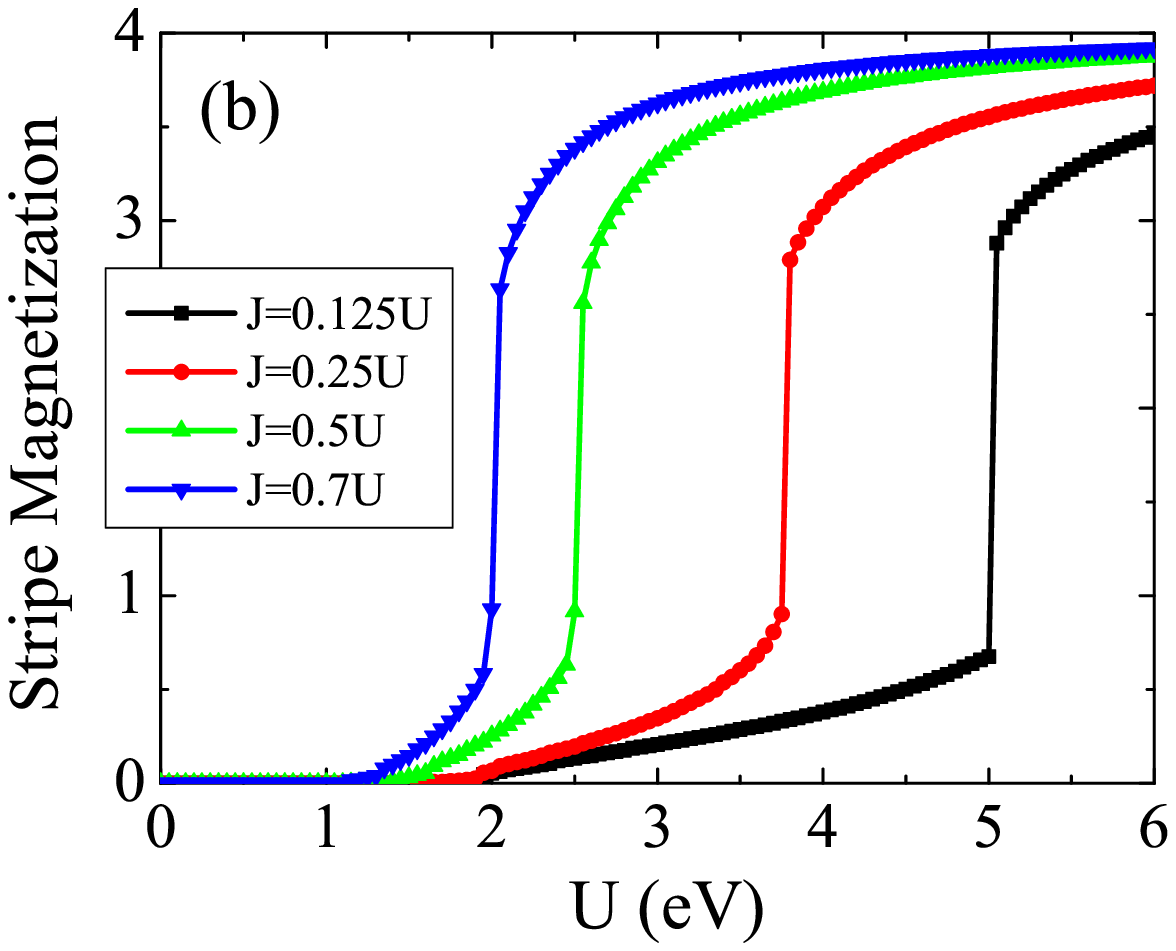}}
\vskip -0.5cm
\caption{(Color online) (a) The mean-field evolution with $U$ of the magnetization of
the state ordered at wavevector $(0,\pi)$, with $J=U/4$. Shown are the two critical
values: one where the magnetization becomes nonzero and a second one where a discontinuous behavior is
concomitant with a metal to insulator transition, as described in the text. (b) The same magnetization
curve as in (a), but for several values of $J/U$. The magnetization shown in (a) and (b) is normalized such that
the maximum value is 4, corresponding to four spin polarized electrons, one per orbital, at each site.}
\vskip -0.7cm
\label{F.MagMF}
\end{center}
\end{figure}

To study in more detail the spin striped ordered state, the
ordering wavevector $\mathbf{q}$ is assumed to be $(0,\pi)$ in the mean-field
approximation. In Fig.~\ref{F.MagMF}(a), the evolution of the
magnetization vs. $U$, at $J$=$U/4$, is shown.
At a critical value $U_{\rm c1}=1.90$~eV, the $(0,\pi)$
order~\cite{noteorder} starts to grow continuously from zero. This ``stripe'' magnetization increases
slowly until it reaches $U_{\rm c2}=3.75$~eV, where it changes
discontinuously, showing the characteristic of a
first-order transition (see discussion for the origin of this transition later in this section). Note that for $U\leqslant U_{\rm c2}$, the
stripe magnetization for the $(0,\pi)$ state is smaller than $1.0$ (with a normalization such
that the maximum possible value is 4.0),
indicating that there is an intermediate $U$ regime that can accommodate
the rather weak striped magnetic order found in the neutron scattering
experiments.\cite{neutrons1,neutrons2}
In Fig.~\ref{F.MagMF}(b), magnetization curves for
the same state at various values of $J$ are presented. Two
transitions for all $J/U$ ratios studied are observed,
similarly as for $J=U/4$ in
(a).

\subsubsection{Band structure and Fermi surfaces}

\begin{figure}[h]
\begin{center}
\centerline{\includegraphics[width=7.5cm,clip,angle=0]{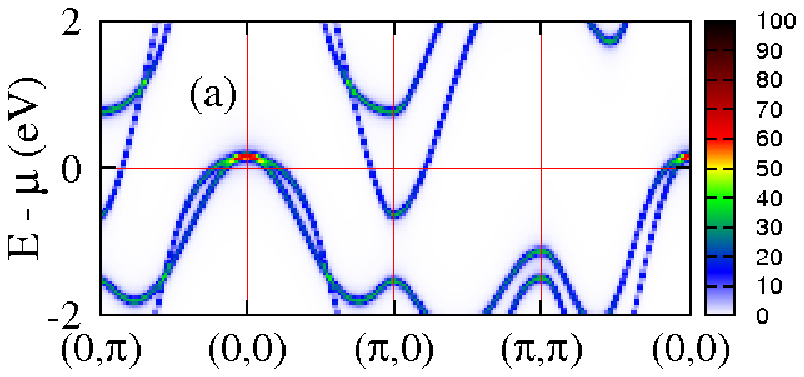}}
\centerline{\includegraphics[width=7.5cm,clip,angle=0]{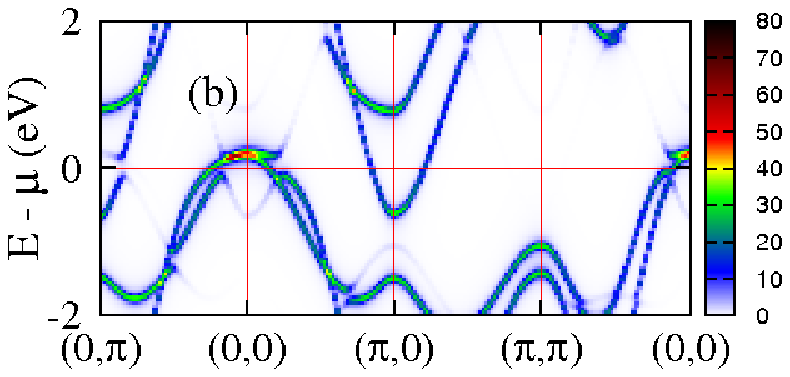}}
\centerline{\includegraphics[width=7.5cm,clip,angle=0]{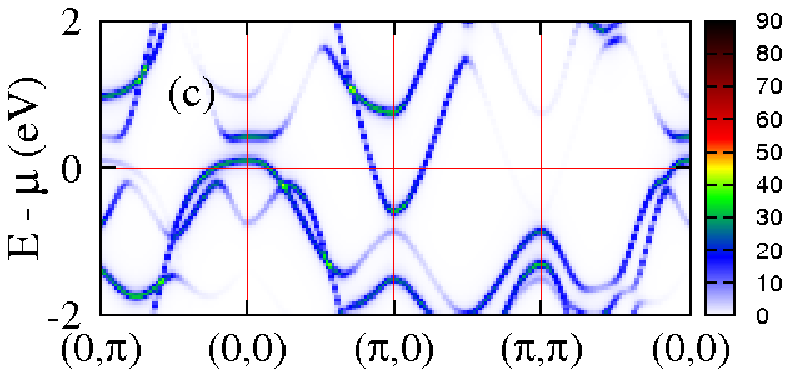}}
\centerline{\includegraphics[width=7.5cm,clip,angle=0]{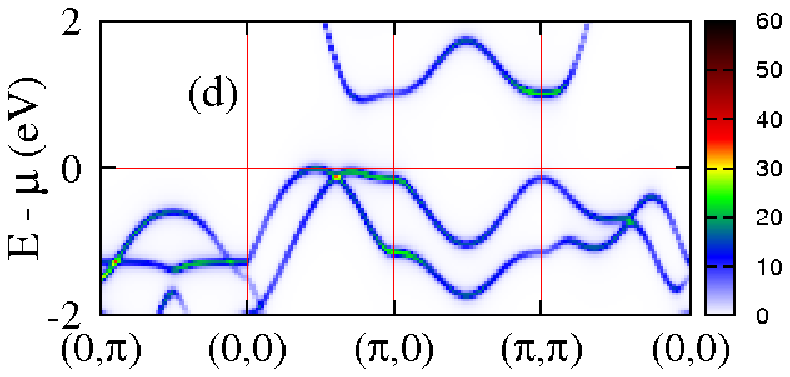}}
%
%
%
\caption{(Color online) Mean-field photoemission band structure of
the spin-striped $(0,\pi)$ state in the
energy window [-2~eV,~2~eV] at $U=1.90$~eV, $U=2.50$~eV, $U=3.20$~eV,
and $U=4.00$~eV, from top to bottom, with $J$=$U/4$. The first $U$ is at the
first transition, and still has the shape of the noninteracting limit.
The second two values of $U$ are in the intermediate coupling
regime, and the presence of states at energy $0$ (location of the chemical
potential) indicate a metallic state. The last coupling, $U=4.00$~eV,
is above the second critical point, and thus already
in the gapped regime. In the two intermediate couplings, 2.50 and 3.20 eV, weak
bands not present in the non-interacting limit (top panel) are revealed. These are the bands
 caused by the striped magnetic order, which should be observable in photoemission experiments.}
\label{F.Band}
\vskip -0.9cm
\end{center}
\end{figure}

\begin{figure}[h]
\begin{center}
\centerline{\includegraphics[width=9cm,clip,angle=0]{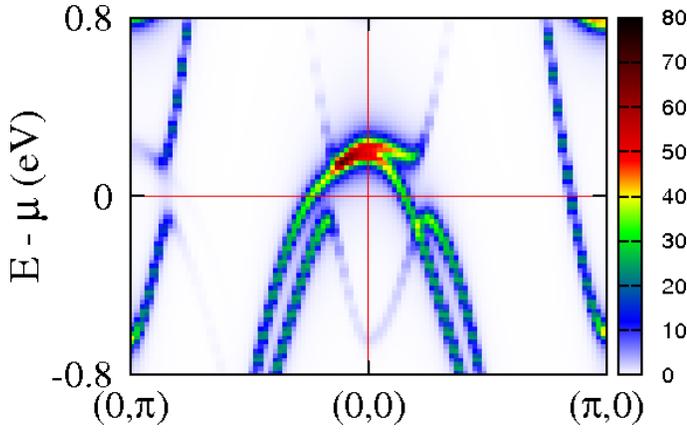}}
\caption{(Color online) Mean-field photoemission band structure of
the spin-striped $(0,\pi)$ state in the
energy window [-0.8~eV,~0.8~eV] at $U=2.50$~eV.
Bands caused by magnetic order
that are not present in the non-magnetic case $U=0.0$ are here shown
in more detail than in Fig.~\ref{F.Band}.
}
\label{F.shadow}
\vskip -0.9cm
\end{center}
\end{figure}

\begin{figure}[h]
\begin{center}
\centerline{\includegraphics[width=6cm,clip,angle=0]{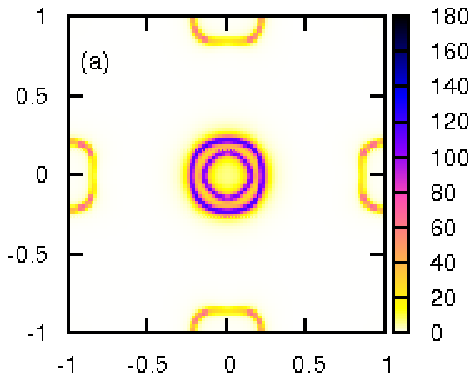}}
\centerline{\includegraphics[width=6cm,clip,angle=0]{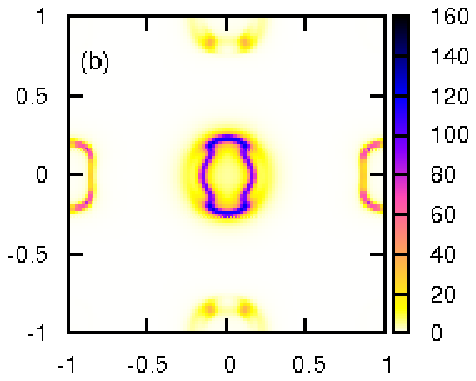}}
\centerline{\includegraphics[width=6cm,clip,angle=0]{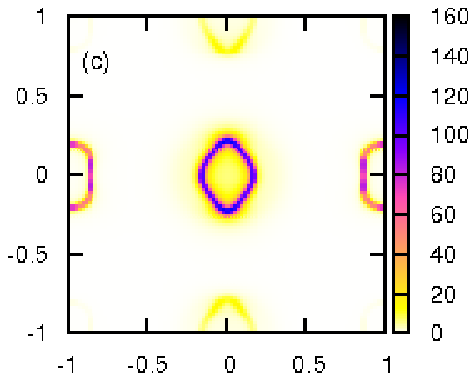}}
\caption{(Color online) Mean-field photoemission Fermi surfaces in
the spin-striped $(0,\pi)$ state at $U=1.90$~eV, $U=2.50$~eV, and
$U=3.20$~eV, from top to bottom. The results are obtained via
$A({\bf k},\omega)$ using a window of 20~meV centered at the Fermi
energy. Shown are results obtained from an equal weight average of
data using $A({\bf k},\omega)$ and $A({\bf p},\omega)$, where ${\bf
k}$=$(k_x,k_y)$ and ${\bf p}$=$(k_x,-k_y)$. By this procedure the
results are properly symmetrized under rotations in the non-magnetic
phase. The anisotropy of the results in the spin-striped
intermediate $U$ region do appear because the spin order breaks
rotational invariance.}\label{F.FS}
\end{center}
\end{figure}

Let us analyze in more detail the spin striped ordered state. Since
varying $J/U$ does not significantly alter the results,
the value $J=U/4$ is adopted in
the rest of the analysis. In Figs.~\ref{F.Band},~\ref{F.shadow}, and \ref{F.FS}
the band structures and Fermi surfaces of this state
at several values of $U$ are extracted from the calculated
mean-field photoemission spectral function data. Both the band structure and
the Fermi surface at the critical point $U_{\rm c1}=1.90$~eV are, of course, identical
to those at $U=0$ in Fig.~\ref{F.BandU0.0}. This
non-interacting electronic description of the system changes gradually upon the establishment of
the striped magnetic order. As discussed in more detail below, gaps open
at particular momenta, while other bands crossing the Fermi surface do not
open a gap. Thus, this is a metallic regime with magnetic order.
Finally, at the magnetization discontinuity a full
gap develops.

\subsubsection{Bands of magnetic origin in the spin striped state}

Due to the off-diagonal term in Eq.~(\ref{E.HMF}), an electron with
momentum $\mathbf{k}$ is coupled to another with momentum
$\mathbf{k+q}$ if their orbital characters are the {\it same}. This generally
leads to avoided level crossings and opens a gap proportional to the
magnetization. If the hole and electron Fermi surfaces can be
connected by the vector $\mathbf{q}$, then a magnetic state with spin
ordering at
$\mathbf{q}$ is stabilized over the non-magnetic state due to the
Fermi surface nesting effect. The magnetic
state gains energy via the opening of a gap near the Fermi level. This
magnetically ordered state significantly changes the band structure
of the non-interacting case, opening gaps and giving
rise to the emergence of new bands of magnetic origin, sometimes called the
``shadow bands'', as shown in Fig.~\ref{F.Band}(b) and (c), and with more detail
for a special case in Fig.~\ref{F.shadow}. Similar issues were discussed in the
context of high temperature superconductors,
where the existence of bands caused by the staggered magnetic order were
extensively studied before.\cite{haas}
In fact, experiments for undoped cuprates revealed a photoemission
spectral function in excellent agreement with theoretical expectations,\cite{wells} i.e.
containing the predicted bands generated by the magnetic order.
Thus, it is to be expected that the spin striped
order of the Fe pnictides should also produce magnetically-induced
bands in the undoped
limit, and even in the doped case if the magnetic correlation length
remains large enough.

\subsubsection{The first order transition at $U_{\rm c2}$}

Regarding the second transition at $U_{\rm c2}$, the
qualitative reason for its presence lies in the incompatibility of the intermediate
coupling metallic state with the large $U$ limit. The mechanism of nesting that
causes the special features of the intermediate $U$ state previously discussed,
including the survival of portions of the Fermi surface,
will lead to an energy that eventually cannot compete with a fully gapped
state at large $U$ at the electronic density considered here, thus
a transition must eventually occur.


But why is the second transition discontinuous? In Fig.~\ref{F.Band}
the coupling between the hole pockets and the electron pocket
at $(0,\pi)$ leads to the distortion of the Fermi surface in the
spin-striped state. Such a coupling between states with a specific
orbital symmetry and momentum also
accounts for the metallic nature of the spin-striped state: the
electron pocket at $(\pi,0)$ is almost undistorted for $U<U_{\rm c2}$.
However, when $U$ is approaching
the second critical value $U_{\rm c2}$, the peak at $(\pi,\pi)$ with occupied states
in the $\alpha_1$ band becomes
energetically closer to
the Fermi level (its energy is increasing with $U$). If the Fermi surface nesting effect could be
neglected at $U_{\rm c2}$, then a smooth behavior would be observed since
the charges could transfer continuously from
the peak at $(\pi,\pi)$, after crossing the Fermi level,
to the peak at $(0,0)$. However, note that
the valley of $\beta_1$ band at $(\pi,0)$
and the peak of the $\alpha_1$ band at $(\pi,\pi)$ have both a
partial $xy$ symmetry. Moreover, they are connected by the
vector $\mathbf{q}=(0,\pi)$. Thus,  it will be
expected that a gap
close to the Fermi level will open to minimize the energy. At
$U_{\rm c2}$, the system
gains maximal energy by opening a finite gap at both $(\pi,0)$
and $(\pi,\pi)$, and
lowering the energy of
the $\alpha$ bands at $(0,0)$
such that they become
fully occupied. This leads to discontinuous changes in the
population of the individual orbitals, producing discontinuous
changes in the orbital magnetizations and, concomitantly,
a finite gap.
Since in the real undoped Fe-pnictide materials the full-gap regime
is not realistic, then we can proceed with
the rest of the analysis below without further consideration of
this discontinuity in the magnetization.

\subsubsection{Anisotropic Fermi surface \\
in the undoped parent compound}

The appearance of spin-striped magnetism with magnetically-induced bands
leads to an {\it anisotropic}
distortion of the Fermi surface in the magnetically ordered state. For instance,
consider the $\mathbf{q}$=$(0,\pi)$ mean-field state. To predict the
pattern of gaps in this state, the information previously discussed in, e.g.,
Fig.~\ref{F.Project}
is important. Through the projected weights, it was observed that the $\beta_1$ electron
pocket centered at $(0,\pi)$ has mainly $yz$+$xy$ symmetry,
while the $(\pi,0)$ electron pocket is mainly $xz$+$xy$.
A dominance of the  $yz$ symmetry is found in the $\alpha_1$ hole pocket along
the $(0,0)\rightarrow(0,\pm \pi)$ directions, and in the $\alpha_2$ hole pocket
along the $(0,0)\rightarrow(\pm \pi,0)$ directions.
%
Then, for the magnetic state with $\mathbf{q}$=$(0,\pi)$,  a gap should open in
the $(0,0)\rightarrow(0,\pm \pi)$ directions for the $\alpha_1$ (inner hole pocket) band and
for the electron pocket at $(0,\pi)$, but the $\alpha_2$ (outer hole pocket) band remains gapless.
In addition, a gap opens in the $(0,0)\rightarrow(\pm \pi,0)$ directions for the
$\alpha_2$ band, but both the inner hole pocket and the $(\pi,0)$
electron pocket remains gapless.
These results  are indeed observed numerically as shown in Figs.~\ref{F.Band}(b,c)

Due to these anisotropic gaps, the Fermi surfaces of the $\alpha$ bands
change their topology from two pockets Fig.~\ref{F.FS}(a) to  four arcs that
are very close to one another, as
shown in Fig.~\ref{F.FS}(b). In between the arcs, there is actually a nonzero but
weak intensity at the location of the original hole pockets.
For LaOFeAs, since a weak striped order
has been observed, the Fermi surface is expected to have a similar shape
as that shown in Fig.~\ref{F.FS}(b). However, topology of the Fermi
surface in this state is sensitive to the value of the coupling $U$:
further increasing this coupling
the four arcs merge into a single pocket Fig.~\ref{F.FS}(c).
Then,
ARPES experiments can provide valuable information about the exact
shape of the Fermi surface and, thus, the interaction strength.

Besides the anisotropic gaps near the
Fermi level, gaps far from the Fermi level also exists (see, for instance,
Fig.~\ref{F.Band}(b)). This cannot occur in a
one-band model because the opening of these gaps would not provide any
energy gain. However, such gaps are possible in more complex multi-orbital
models: since the off-diagonal term in Eq.~(\ref{E.HMF}) depends on the
magnetization that is contributed by each orbital, once the magnetic state
is stabilized by the opening of a gap near the Fermi level, the
off-diagonal term becomes non-zero even for the bands far from the Fermi
level.

\begin{figure}
\begin{center}
\vskip -0.3cm
\centerline{\includegraphics[width=9cm,clip,angle=0]{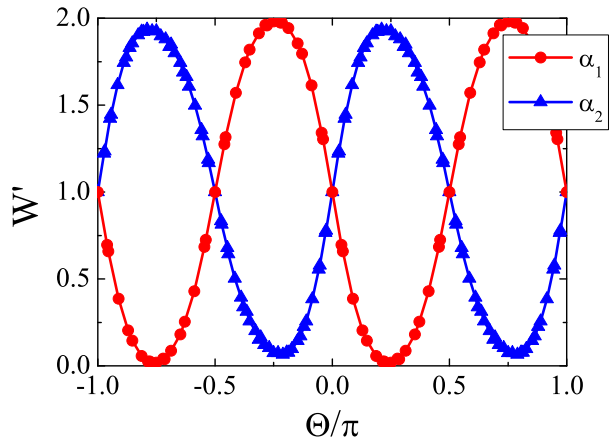}}
\vskip -0.5cm
%
\caption{(Color online) The angular resolved weight $W'$ at $U$=0
obtained using
Eq.~(\ref{E.Wp}) for the two $\alpha$ hole-like pockets. The
definition of $\Theta$ is the same as in
Fig.~\ref{F.Project}.}
\vskip -0.7cm
\label{F.Wp}
\end{center}
\end{figure}

A rather surprising result is that the intensities of the spectral
function of the two $\alpha$ pockets display an anisotropy even
in the non-magnetic phase (not shown).
This is puzzling since from Fig.~\ref{F.Project}, it can be observed that $\sum_{\mu}
W_{\mu,\lambda} (\mathbf{k})= 1$ for any $\mathbf{k}$. Naively, this
would lead to an isotropic $A(\mathbf{k},\omega)$. However, from
Eq.~(\ref{E.Akw}), $A(\mathbf{k},\omega)= \sum_{\lambda}
\delta(\omega-\rho_\lambda) W'_{\lambda}(\mathbf{k})$, where
\begin{eqnarray}\label{E.Wp}
W'_{\lambda}(\mathbf{k}) =
\sum_{\sigma}|\sum_{\mu}U_{\mathbf{k},\mu,\sigma;\lambda}|^2.
\end{eqnarray}
Here, note that an anisotropy could arise from the interference between
different orbitals. To observe this more explicitly,  $W'_\lambda$ for
the two $\alpha$ hole pockets at $U=0$ is shown
in Fig.~\ref{F.Wp}. These functions are not
constant but oscillate between $0$ and $2$. From Fig.~\ref{F.Wp}, it is
observed that for the $\alpha_1$ band $W'$ reaches its maximum at, e.g.,
$-\pi/4$, corresponding to the $(0,0)\rightarrow(-\pi,\pi)$
direction, while for the $\alpha_2$ band, the maximum of $W'$ is at, e.g.,
$\pi/4$, i.e., in the $(0,0)\rightarrow(\pi,\pi)$ direction.
Such an anisotropy in $A(\mathbf{k},\omega)$ is not a
consequence of the mean-field approximation, but is related to the
multi-band nature of the model itself.
Interestingly, such an anisotropy in the topology of the Fermi surface
in the $A(\mathbf{k},\omega)$ data may account for the anisotropic
features of the hole pocket in the recent ARPES experiment on
LaOFeP,\cite{LaOFeP_ARPES} where a long-ranged striped magnetic order is
absent.\cite{LaOFeP_SDW}
However,
in Fig.~\ref{F.FS} this problem is avoided by a symmetrization procedure (see
caption of Fig.~\ref{F.FS}) that restores rotational invariance to the non-magnetic
state. Thus, the anisotropy of the important intermediate $U$ regime fully originates
in the lack of rotational invariance of the spin striped state with a wavevector
$(0,\pi)$ or $(\pi,0)$.

\subsubsection{Metallic magnetically ordered phase at intermediate couplings and
existence of a pseudogap}

\begin{figure}[h]
\begin{center}
\vskip -0.5cm
\centerline{\includegraphics[width=9.0cm,clip,angle=0]{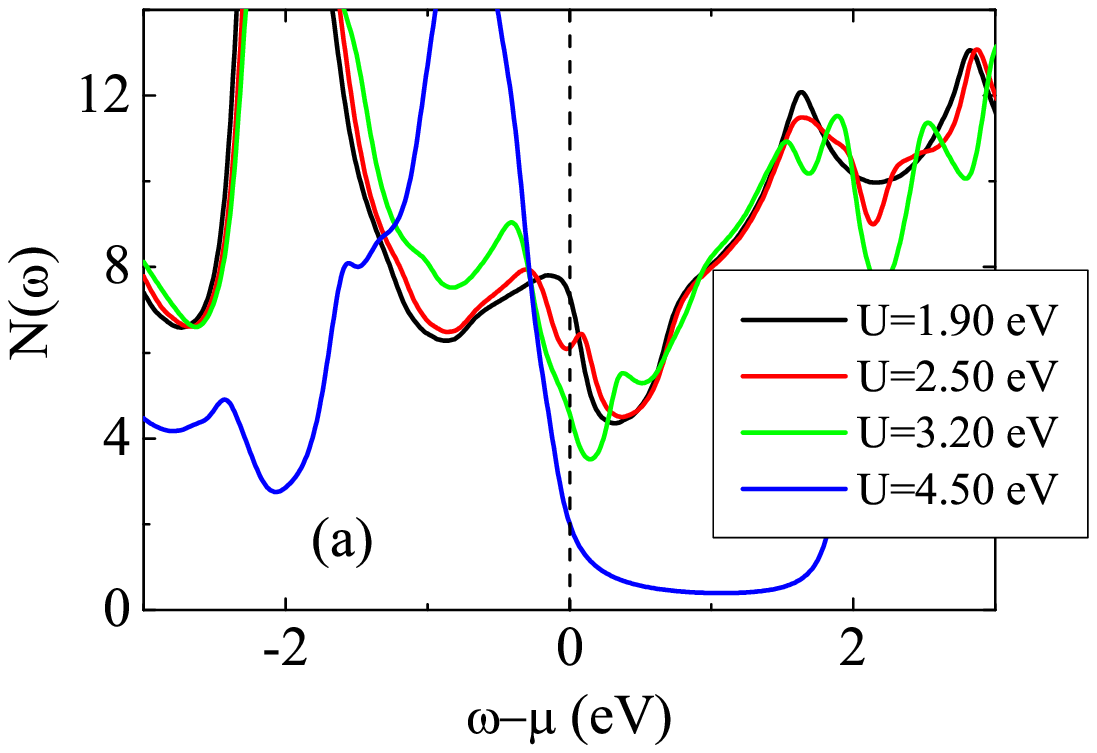}}
\vskip -0.7cm
\centerline{\includegraphics[width=9.0cm,clip,angle=0]{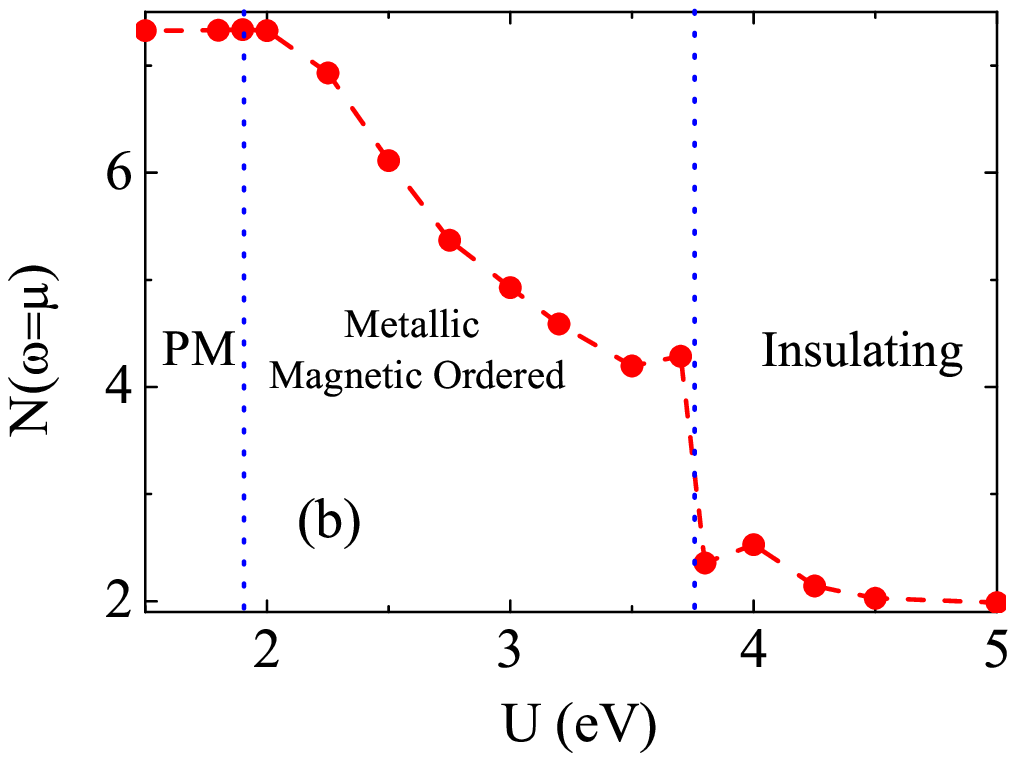}}
\vskip -0.5cm
%
\caption{(Color online) (a) Density of states at various values of
$U$ showing the development of a pseudogap at the chemical potential
in the spin-striped $(0,\pi)$ state. (b) The evolution of $N(\omega)$, the DOS at the
chemical potential, indicates three well separated regions.
The non-zero values of $N(\omega=\mu)$ in the
insulating phase at $U>U_{\rm c2}$ arises from the finite broadening of
the raw DOS numerical data.}
\vskip -0.5cm
\label{F.DOSMF}
\end{center}
\end{figure}

As already remarked,
Fig.~\ref{F.Band} indicates that for moderate $U$ the
magnetically ordered system is still gapless, i.e. it is in a metallic
phase with a finite Fermi surface due to the phenomenon
of ``band overlapping''
described in the Introduction. This intermediate state is also revealed via the
evolution of the DOS in Fig.~\ref{F.DOSMF}. As displayed in
Fig.~\ref{F.DOSMF}(a), a pseudogap in the DOS near the chemical
potential exists in the regime $U_{\rm c1}<U<U_{\rm c2}$. If the availability
of states at the Fermi level is assumed to be directly related with transport
properties, a DOS pseudogap suggests bad-metallic
characteristics in the intermediate $U$ regime. The pseudogap turns into a hard gap
at $U>U_{\rm c2}$, where the system becomes an insulator.
In Fig.~\ref{F.DOSMF}(b) the value of the DOS at the
chemical potential vs. $U$ is plotted.
Two transitions can be easily identified. For
$U<U_{\rm c1}$, $N(\mu)$ is a constant. It decreases continuously
for $U>U_{\rm c1}$, indicating a second-order transition from the
paramagnetic metallic phase to the metallic phase with striped
magnetic order. At $U>U_{\rm c2}$, it drops abruptly to a small value,
corresponding to the formation of a full gap in the insulating
phase (the finite DOS at large $U$ is simply caused by the
artificial broadening of delta functions to plot results).

\subsubsection{Stability of the striped phase}

\begin{figure}[h]
\begin{center}
\vskip -0.3cm
\centerline{\includegraphics[width=9cm,clip,angle=0]{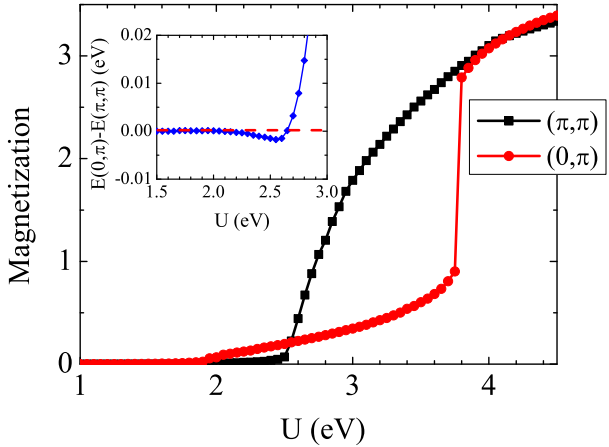}}
\vskip -0.4cm
%
\caption{(Color online) Main panel: the evolution of the magnetization
with $U$ for $\mathbf{q}$=$(0,\pi)$ and $(\pi,\pi)$. Inset: Energy
difference between the two states of the main panel.}
\vskip -0.5cm
\label{F.MagMF2}
\end{center}
\end{figure}

Consider now the stability of the striped
spin-order state by analyzing two possible magnetically ordered
states, one with $\mathbf{q}$=$(0,\pi)$ and another one with
$(\pi,\pi)$. The $U$
dependence of the magnetization and the energy difference
between these two states are presented in Fig.~\ref{F.MagMF2}. The
$(\pi,\pi)$ order appears at $U\approx2.5$~eV, higher than the $U_{\rm c1}$
for the $(0,\pi)$ state. However, with increasing $U$, the $(\pi,\pi)$
staggered magnetization increases much
faster than for the striped state. As shown in the inset
of Fig.~\ref{F.MagMF2}, the $(\pi,\pi)$ state becomes more stable than the
$(0,\pi)$ state for $U\geqslant2.65$~eV. We have analyzed several other sets
of hopping parameters and ratios $J/U$ and the results are qualitatively the same:
the spin stripes are stable, but in a narrow region of couplings.

Hence, in our model the stripe-order ground state becomes unstable to
another antiferromagnetic state with $\mathbf{q}$=$(\pi,\pi)$ at
$U\geqslant2.65$~eV through a first-order transition. This is not too
surprising considering the values of the hopping parameters: the NN
and NNN hoppings are similar in magnitude. Thus, there is a competition
between different spin tendencies, as it occurs at intermediate couplings
in Heisenberg spin systems with
NN and NNN terms.
%
However, the neutron scattering experiments
show that the ground state of several undoped Fe-pnictides presents {\it only} the
striped spin order. To understand this predominance of one state
over the other, note that it has been argued that the emergence of the striped state is
closely related to a structural phase
transition~\cite{neutrons1,neutrons2} from space group $P4/nmm$ to
$P112/n$. The lattice distortion at the
transition breaks the four-fold rotational symmetry and lifts the
degeneracy of the $xz$  and $yz$ orbitals. Thus, according to recent
calculations that incorporate the lattice distortion, the system prefers the
$(0,\pi)$ state over the $(\pi,\pi)$ state.\cite{yildirim}
Since the effects of lattice distortions are not
included in our model, then the $(\pi,\pi)$ spin-ordered state
apparently strongly competes with the spin stripes, but this is misleading
and caused by the absence of lattice energetic considerations.

\begin{figure}[h]
\begin{center}
\centerline{\includegraphics[width=8.5cm,clip,angle=0]{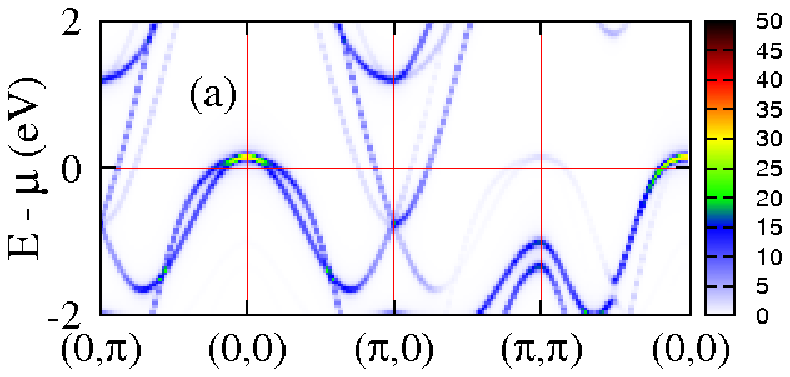}}
\centerline{\includegraphics[width=8.5cm,clip,angle=0]{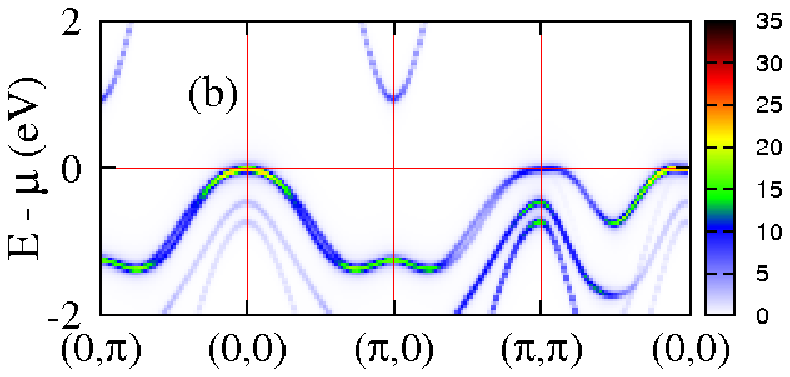}}
%
\caption{(Color online) Mean-field band structures of the spin-ordered
$(\pi,\pi)$ state in the
energy window [-2~eV,~2~eV]. The upper panel is obtained
at $U=2.75$~eV where band overlap indicates a metallic state, even if the
order parameter is nonzero.  The lower panel is at
$U=4.50$~eV where the gap is fully developed.}
\vskip -0.9cm
\label{F.BandG}
\end{center}
\end{figure}

This discussion leads us to believe that simply analyzing the
spin-stripe state found in the mean-field approximation should be sufficient
to understand some of the electronic properties of the real system. However,
it is remarkable that even in the competing $(\pi,\pi)$
ordered state there are also two magnetically-ordered phases: a metallic phase for
moderate $U$ values, and an insulating phase with a finite gap at
larger $U$ values (see Fig.~\ref{F.BandG}).
This qualitatively agrees with the previous analysis
for the striped ordered phase. This suggests that the existence of
a metallic magnetically ordered phase at moderate $U$ is
an intrinsic property of the multi-orbital Hubbard model, treated in the mean-field
approximation, that is robust
varying the interactions, as discussed in more detail in the next subsection.

\subsubsection{Mean-field results for other models \\
with several active orbitals}

We have applied the mean-field technique to several other
multi-orbital models such as a five-orbital model,\cite{kuroki} another
four-orbital model,\cite{korshunov} and an effective
three-orbital model including the $xz$, $yz$, and $xy$ orbitals.\cite{plee}  For all
these multi-band models a Coulombic interaction term similar to
Eq.~(\ref{E.Hint}) has been used. A robust conclusion of our mean-field analysis is that a
transition from a paramagnetic metal to a metallic striped
spin-order state is found in all the models considered here.

\begin{figure}[h]
\begin{center}
\vskip -0.7cm
\centerline{\includegraphics[width=9cm,clip,angle=0]{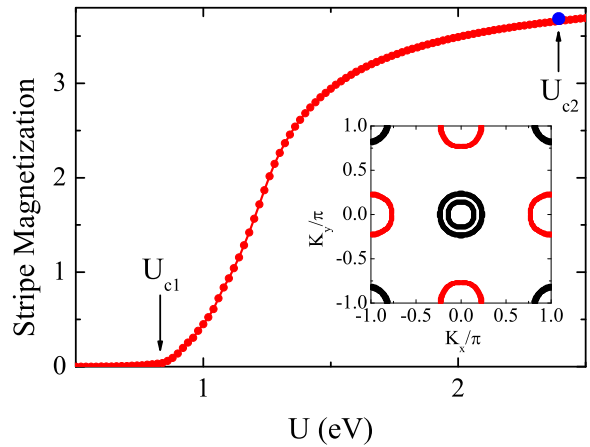}}
\vskip -0.5cm
%
\caption{(Color online) Main panel: the evolution of the magnetization
for the state with $\mathbf{q}$=$(0,\pi)$ in a five-orbital
model,\cite{kuroki} at electron filling $n=6.0$. Inset: the Fermi surface at $U=0$.}
\vskip -0.8cm
\label{F.Kurokin6.2}
\end{center}
\end{figure}

Let us discuss the results for the five-orbital model.\cite{kuroki}
This model does not give precisely the correct Fermi surface
topology in the undoped case, since it contains a pocket at $(\pi,\pi)$.
This problem can be fixed by slightly modifying the electronic concentration to, e.g., $n=6.2$.
However, in our study we will consider $n=6.0$ for consistency
with the rest of the analysis.
Within the mean-field approximation, the magnetization at
$n=6.0$ is shown in Fig.~\ref{F.Kurokin6.2}. In this case there is a broad region
of metallicity, and the second critical Hubbard coupling (also indicated)
does not involve a discontinuity.
The Fermi
surface in the inset shows pockets at $(0,0)$, $(\pi,0)$, $(0,\pi)$, and $(\pi,\pi)$.
We conclude that a variety of multi-orbital models show similar
features as the four-orbital model analyzed before, particularly
regarding an intermediate metallic magnetic phase.

\section{Results for the two-orbital model}\label{sec:twoorbs}

The complexity of four- and five-orbital Hamiltonians leads to very
large Hilbert spaces even for small clusters and, as a consequence,
it is not possible to compare the mean field results against
Exact Diagonalization (ED)\cite{RMP} or Variational Cluster
Approach (VCA)\cite{Aic03,Pot03} results. For this reason
the two-orbital model, which was recently studied
with ED and VCA methods,\cite{daghofer} is here revisited to assess the validity
of the mean-field approximation. Our conclusion is that all
three techniques lead to similar results for the two-orbital model,
lending support to  our claim for the
existence of an intermediate-coupling metallic
and magnetic state.

The two-orbital model has been widely analyzed in recent literature
and its derivation and
other properties will not be repeated here. It is given by
\begin{equation}
H_{2o}=H_{\rm K}+H_{\rm int},
\label{1}
\end{equation}
where the kinetic energy $H_\textrm{K}$
in real space is:\cite{daghofer,scalapino}
\begin{equation}\begin{split}
H_{\rm K}&=-t_1\sum_{{\bf i},\sigma}(d^{\dagger}_{{\bf i},x,\sigma}
d_{{\bf i}+\hat y,x,\sigma}+d^{\dagger}_{{\bf i},y,\sigma}
d_{{\bf i}+\hat x,y,\sigma}+h.c.)\\
&\quad -t_2\sum_{{\bf i},\sigma}(d^{\dagger}_{{\bf i},x,\sigma}
d_{{\bf i}+\hat x,x,\sigma}+d^{\dagger}_{{\bf i},y,\sigma}
d_{{\bf i}+\hat y,y,\sigma}+h.c.)\\
&\quad
-t_3\sum_{{\bf i},\hat\mu,\hat\nu,\sigma}(d^{\dagger}_{{\bf i},x,\sigma}
d_{{\bf i}+\hat\mu+\hat\nu,x,\sigma}+d^{\dagger}_{{\bf i},y,\sigma}
d_{{\bf i}+\hat\mu+\hat\nu,y,\sigma}+h.c.)\\
&\quad+t_4\sum_{{\bf i},\sigma}(d^{\dagger}_{{\bf i},x,\sigma}
d_{{\bf i}+\hat x+\hat y,y,\sigma}+d^{\dagger}_{{\bf i},y,\sigma}
d_{{\bf i}+\hat x+\hat y,x,\sigma}+h.c.)\\
&\quad-t_4\sum_{{\bf i},\sigma}(d^{\dagger}_{{\bf i},x,\sigma}
d_{{\bf i}+\hat x-\hat y,y,\sigma}+d^{\dagger}_{{\bf i},y,\sigma}
d_{{\bf i}+\hat x-\hat y,x,\sigma}+h.c.)\\
&\quad-\mu\sum_{\bf i}(n_{{\bf i},x}+n_{{\bf i},y}).
\end{split}\end{equation}
The form of $H_{\rm K}$ in momentum space was provided in
Refs.~\onlinecite{daghofer} and \onlinecite{scalapino} and is given by
Eqs.~(\ref{eq:t11})-(\ref{eq:t12}) of the four-orbital model.
The Coulomb interaction terms are
\begin{equation}\label{eq:Hint2}\begin{split}
H_{\rm int}&=
U\sum_{{\bf i},\alpha}n_{{\bf i},\alpha,\uparrow}n_{{\bf i},
\alpha,\downarrow}
+(U'-J/2)\sum_{{\bf i}}n_{{\bf i},x}n_{{\bf i},y}\\
&\quad-2J\sum_{\bf i}{\bf S}_{{\bf i},x}\cdot{\bf S}_{{\bf i},y}\\
&\quad+J\sum_{{\bf i}}(d^{\dagger}_{{\bf i},x,\uparrow}
d^{\dagger}_{{\bf i},x,\downarrow}d_{{\bf i},y,\downarrow}
d_{{\bf i},y,\uparrow+h.c.)},
\end{split}\end{equation}
where the notation is the same as for the case of the four-orbital model
but with $\alpha=x,y$ here denoting the orbitals $xz$ and $yz$.
The index $\hat\mu$ is a unit vector linking NN sites and takes the values
$\hat x$ or $\hat y$.
%
%
$\mu$ is the chemical potential. As for the case of four orbitals, the relation
$U'$=$U-2J$ originating in
rotational invariance\cite{RMP01} was used. In the
last term in Eq.~(\ref{eq:Hint2}), the same rotational invariance also establishes that
the pair hopping coupling $J'$ must be equal to $J$.
As before,
the hoppings are determined from the orbital integral overlaps within the
SK formalism or from LDA band
dispersion fittings.\cite{daghofer,scalapino}
Most of the properties of this two-orbital model are
formally similar to those of the four-orbital model and,
as a consequence, several details of the analysis of the previous section
do not need to be repeated here.

\subsection{Mean field approximation}

Following the same procedure used for the four-orbital model, here
we consider only the mean field values that are
diagonal with respect to the Fe site, orbital, and spin labels such that
\begin{equation}
\langle d^{\dagger}_{{\bf l},\alpha,\sigma}d_{{\bf l'},\alpha',\sigma'}\rangle=(n_{\alpha}
+{{\sigma}\over{2}}\cos({\bf q.r}_{\bf l})m_{\alpha})\delta_{\bf ll'}\delta_{\alpha\alpha'}
\delta_{\sigma\sigma'}.
\label{4}
\end{equation}

After introducing Eq.~(\ref{4}) to decouple the four-fermion interactions
in Eq.~(\ref{eq:Hint2}), and then transforming into momentum space, we obtain:
\begin{equation}\begin{split}\label{mf2}
H^{\rm MF}_{\rm int}&=
-UN\sum_{\alpha}(n^2_{\alpha}-{1\over{4}}m^2_{\alpha})-4(U'-{J\over{2}})Nn_xn_y\\
&\quad +\frac{JNm_xm_y}{2}+\sum_{{\bf k},\sigma}\left(Un_x+2(U'-{J\over{2}})n_y\right)n_{{\bf k}x\sigma}\\
&\quad+\sum_{{\bf k},\sigma}\left(Un_y+2(U'-{J\over{2}})n_x\right)n_{{\bf k}y\sigma}\\
&\quad-{1\over{4}}\sum_{{\bf k}\sigma}(Um_x+Jm_y)(d^{\dagger}_{{\bf k}x\sigma}d_{{\bf k+q}x\sigma}
+d^{\dagger}_{{\bf k+q}x\sigma}d_{{\bf k}x\sigma})\\
&\quad-{1\over{4}}\sum_{{\bf k}\sigma}(Um_y+Jm_x)(d^{\dagger}_{{\bf k}y\sigma}d_{{\bf k+q}y\sigma}
+d^{\dagger}_{{\bf k+q}y\sigma}d_{{\bf k}y\sigma}).
\end{split}\end{equation}

The four mean-field parameters $n_x$,
$n_y$, $m_x$, and $m_y$ are determined in the usual way by minimizing the
energy and by requesting that the system be half-filled. The half-filling
condition determines that $n_x=n_y=0.5$ while the values of $m_x$ and $m_y$
are a function of $U$ and $J$. We have observed that varying $U$, and
for fixed $J$, $m_{\alpha}$ becomes non-zero at a critical value $U=U_{\rm c1}$ where
a gap opens separating
the ``valence'' and ``conduction'' bands at particular momenta but, overall,
there is still an overlap in energy of some bands and the
chemical potential crosses both of them. Thus, the system has developed magnetic order
but it is still a metal. However, the bands no longer overlap when $U>U_{\rm c2}$ and, thus,
a metal insulator transition occurs. These results are similar to those obtained
using more orbitals
in the model.

\begin{figure}
\centerline{\includegraphics[width=10cm,clip,angle=0]{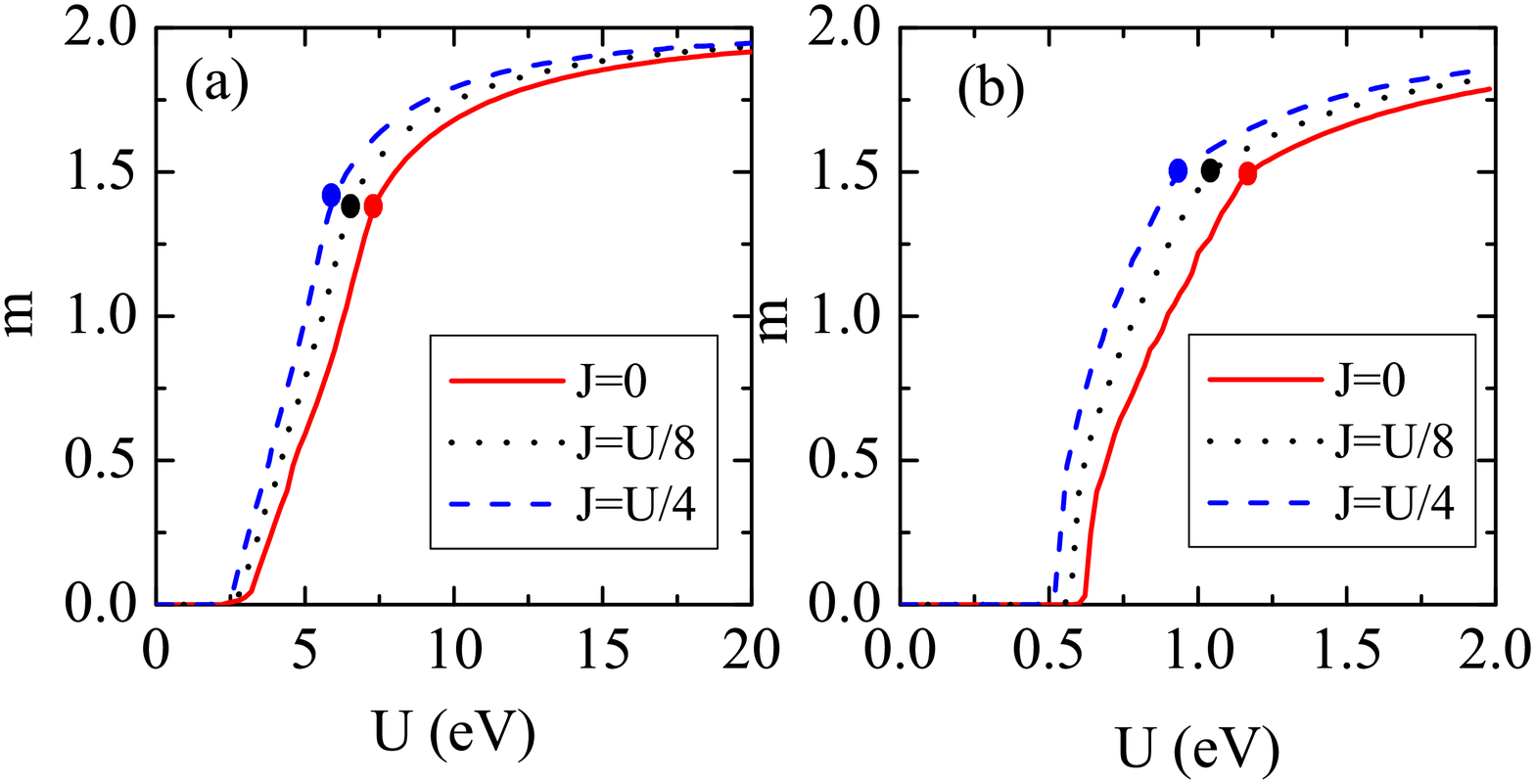}}
\vskip -0.5cm
\caption{(Color online) Two-orbital model mean-field calculated spin-stripe magnetization $m$ vs.
$U$. Panel (a) is for the LDA fitted hoppings,\cite{scalapino} while panel (b) is for the
SK hoppings.\cite{daghofer}  The solid red line is for
$J$=$0$. With a blue dashed line the results at $J$=$U/4$ are indicated. The dotted black line
denotes results for $J$=$U/8$. The large dot in each curve indicates the value of $m$ at
$U_{\rm c2}$
where the metal to insulator transition occurs (see text).
Note the absence of a first-order transition at $U_{\rm c2}$
as in the four-orbital model.
}
\label{kien1}
\end{figure}

In Fig.~\ref{kien1}, the stripe magnetization $m$=$m_x+m_y$ vs. $U$ is shown
for $J$=$0$, $U/4$, and $U/8$. Panel (a) contains the results for the LDA
fitted
hoppings\cite{scalapino} while panel (b) is for the SK hoppings\cite{daghofer} with
$pd\sigma=-0.2$. The critical coupling $U_{\rm c1}$ is the value of $U$ where
$m$ becomes different from zero. There is a second critical coupling
$U_{\rm c2}$ that is obtained by monitoring the density of states, and it separates a metallic from an
insulating regime. For completeness, this second coupling  is indicated with a full circle for each case.
In panel (b), the shape of the curves  is the same for the three values of $J$ investigated, and the actual
value of $U_{\rm c1}$ mildly depends on $J$.

\begin{figure}
\centerline{\includegraphics[width=10cm,clip,angle=0]{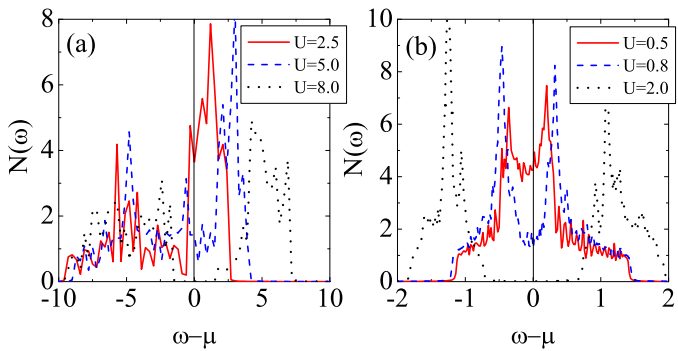}}
\vskip -0.5cm
\caption{(Color online) Two-orbital model mean-field calculated density of states
(a) for the LDA fitted hoppings;\cite{scalapino} (b) for the
SK hoppings.\cite{daghofer} Panel (a): the solid red line is for
$U$=$2.5$, blue dashed line for $U=5.0$, and dotted
black line for $U=8.0$. Panel (b): red is for $U$=0.5,
blue is for $U$=0.8, and black is for $U$=2.0.
$J$=$U/4$ is used in both panels.}
\label{kien2}
\end{figure}

The mean-field density of states (DOS) $N(\omega)$
is presented in Fig.~\ref{kien2} for some values of $U$ and $J$=$U/4$, and the
two sets of hoppings considered here.
The solid curve in both panels
shows results for $U<U_{\rm c1}$. In this case the system is metallic.
The dashed curve displays the DOS
for a value of $U$  in between the two critical points. Although the DOS varies continuously as $U$ increases,
it is clear that this regime is qualitatively different: A deep
pseudogap has developed at the chemical potential.
The system is still metallic in this regime, although
likely with ``bad metal'' characteristics.
Finally, the dotted curve shows the DOS for $U>U_{\rm c2}$. In this
case, there is a gap at the
chemical potential and the system has become
insulating. Interestingly, the transition at $U_{\rm c2}$ is not
first order for the two-orbital model, in contrast to the case with
four orbitals.

\subsection{Mean field results for the bands and Fermi surface}

\begin{figure}
\centerline{\includegraphics[width=9cm,clip,angle=0]{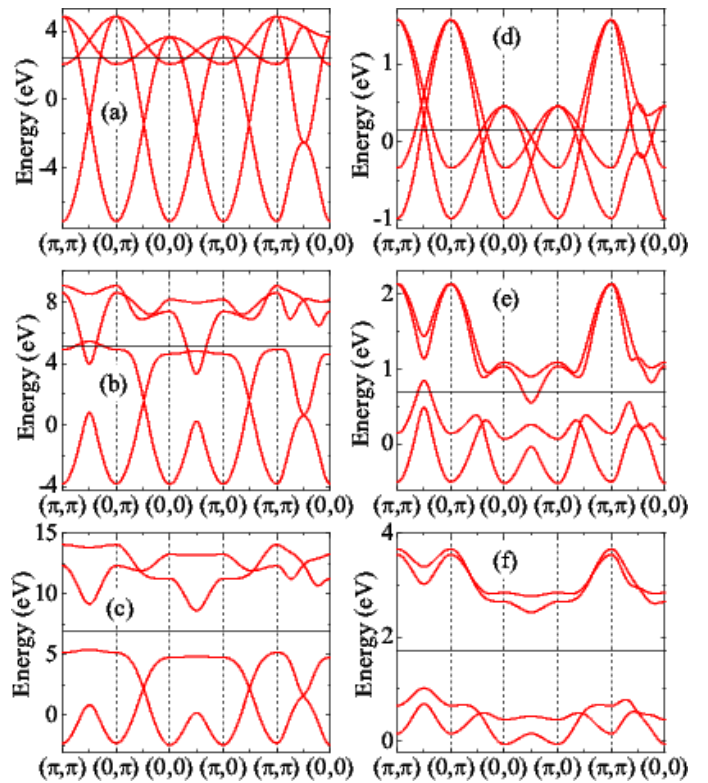}}
\caption{(Color online) Two-orbital model mean-field
band structure along high symmetry directions in the extended FBZ.
The panels on the left column show results for the LDA fitted
hoppings,\cite{scalapino} while results on the right are for the SK
hoppings.\cite{daghofer} (a) $U$=$2.5$, (b) $U$=$5.0$, (c)
$U$=$8.0$, (d) $U$=$0.5$, (e) $U$=$0.8$, (f) $U$=$2.0$. In all
cases, $J$=$U/4$ and the magnetic order wavevector is $(\pi,0)$. }
\vskip -0.1cm \label{kien3}
\end{figure}


\begin{figure}[thbp]
\begin{center}
\includegraphics[width=5.8cm,clip,angle=0]{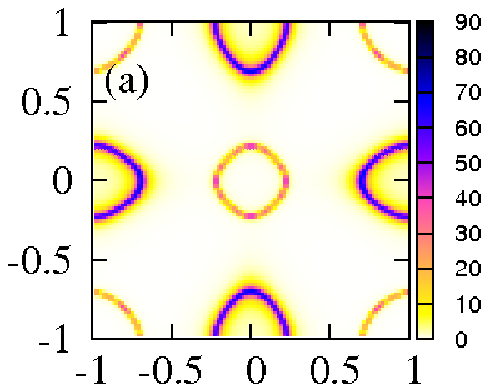}
\includegraphics[width=5.8cm,clip,angle=0]{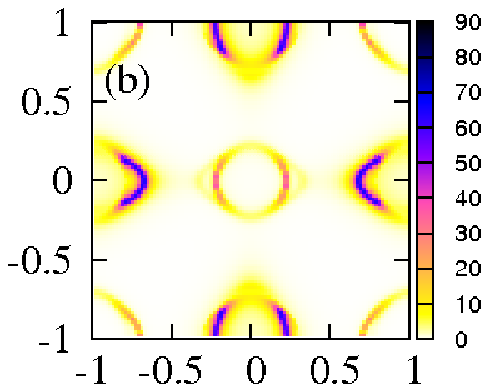}
\includegraphics[width=6cm,clip,angle=0]{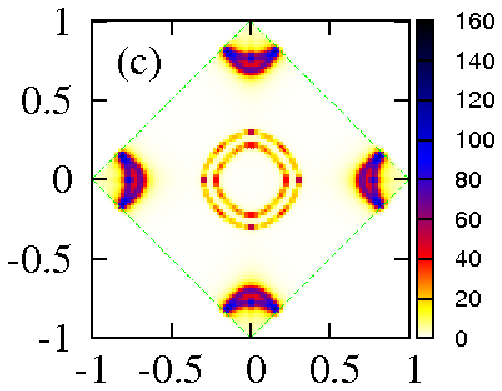}
\includegraphics[width=6cm,clip,angle=0]{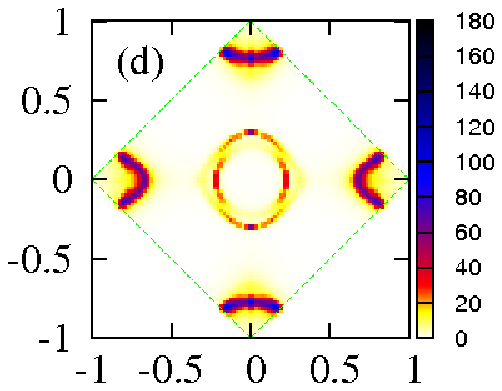}
\caption{(Color online) Mean-field photoemission Fermi surface
for the LDA fitted hoppings\cite{scalapino}
in the unfolded FBZ for (a) $U=1$, and (b) $U=3$, and in the folded FBZ
for (c) $U=1$, and (d) $U=3$. The ratio $J=U/4$ was used, and the
magnetic order wavevector is $(\pi,0)$. These results
were obtained via $A({\bf k},\omega)$ using the same
symmetrization procedure and energy window centered at the Fermi energy as in Fig.~\ref{F.FS}.
}
\vskip -0.5cm
\label{kien4}
\end{center}
\end{figure}

The mean-field band structure obtained by solving the mean-field
self-consistent equations is shown along high-symmetry directions in the Brillouin zone
in Fig.~\ref{kien3}. The panels on the left (right) column correspond to LDA fitted
(SK) hoppings.
The top row shows the band dispersion for $U<U_{\rm c1}$. The corresponding Fermi surfaced (FS) in the
extended and reduced Brillouin
zones (BZ) are shown in Figs.~\ref{kien4}~(a) and (c), and they agree with previous discussions
for the two-orbital model.\cite{daghofer}



The second row of panels in
Fig.~\ref{kien3} shows the band dispersion in the interesting regime
$U_{\rm c1}<U<U_{\rm c2}$. It can be observed that gaps have opened
along, e.g., the direction $(0,0)$-$(0,\pi)$ but not along
other high-symmetry directions. The partial gaps remove portions
of the original FS and
produce the $A({\bf k},\omega)$ disconnected features (arcs) at $\Gamma$
shown in panel (b) of Fig.~\ref{kien4}.
Note that in Fig.~\ref{F.FS}~(b), the results for the four-orbital
model also contained arcs, but they appeared in a more symmetric manner, namely with four arcs surrounding
the $\Gamma$ point. Two of those arcs were located in the inner hole pocket and two in the outer hole
pocket, while in Fig.~\ref{kien4}~(b) there are only two arcs around the $\Gamma$ point. However, as
often remarked in the context of the two-orbital model, to compare with either experiments or with results
of models with more orbitals, it is crucial to fold the results since by this mechanism the hole pocket
at $M$ becomes the outer hole pocket at $\Gamma$. Following this procedure,
in Fig.~\ref{kien4}~(d) the folded results
are shown and now there are {\it four arcs} around the $\Gamma$ point in good agreement with the
results of the four-orbital model. We conclude that in the interesting intermediate coupling regime,
both models give similar results upon folding of the extended Brillouin zone.


The last row of panels in Fig.~\ref{kien3} shows the band dispersion for $U>U_{\rm c2}$. Now the upper and lower bands no longer
overlap. The gap is complete and there is no FS.
The system has become an insulator, as it can be seen in the DOS shown in
Fig.~\ref{kien2}.

\begin{figure}[thbp]
\begin{center}
\includegraphics[width=8cm,clip,angle=0]{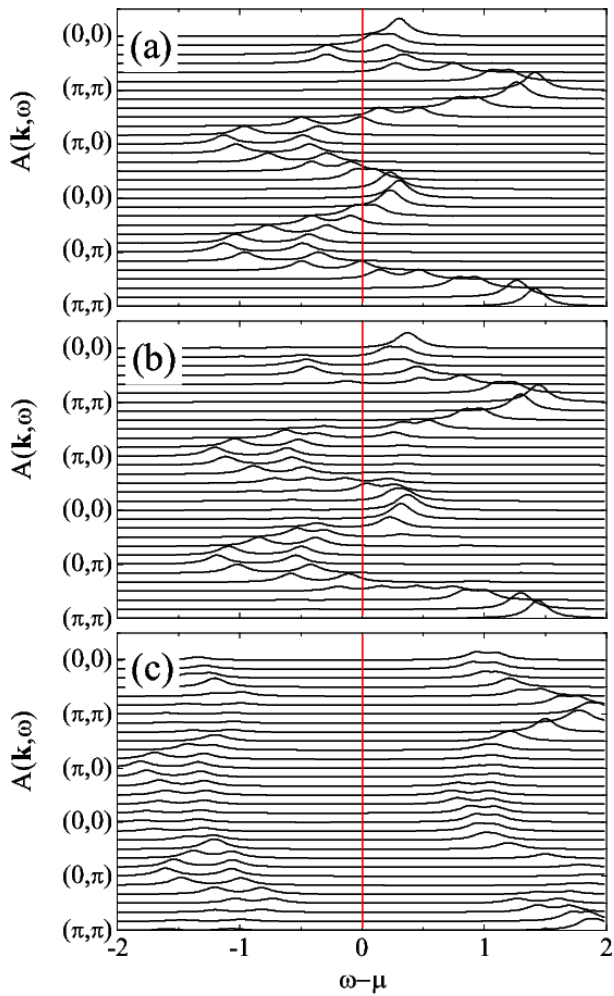}
\caption{(Color online) Two-orbital model mean-field spectral function along high symmetry
directions
in the extended FBZ, using SK hoppings.\cite{daghofer}
(a) $U$=$0.5$, (b) $U$=$0.8$, (c) $U$=$2.0$. The Hund coupling is fixed to $J$=$U/4$ and
the magnetic order wavevector is $(\pi,0)$.
}
\vskip -0.5cm
\label{kien5}
\end{center}
\end{figure}

\begin{figure}[thbp]
\begin{center}
\includegraphics
[width=8cm,angle=0,bbllx=14pt,bblly=14pt,bburx=241pt,bbury=383pt]{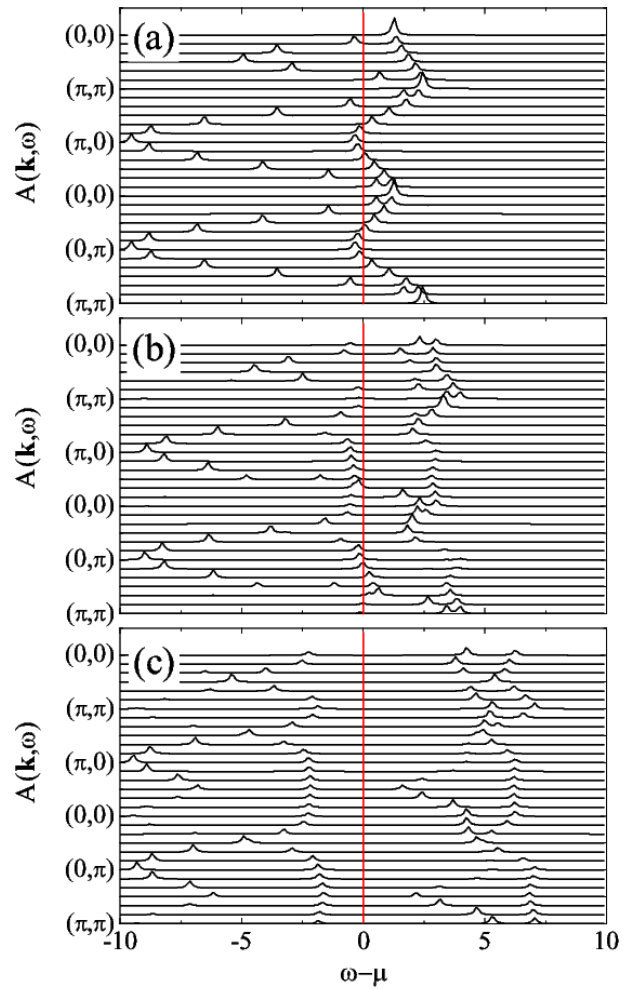}
\caption{(Color online) Two-orbital model mean-field spectral function along high symmetry
directions
in the extended FBZ for the LDA fitted hoppings.\cite{scalapino}
The couplings used are (a) $U$=$2.5$, (b) $U$=$5$, (c) $U$=$8$. In all cases
$J$=$U/4$, and the magnetic order wavevector is $(\pi,0)$.
}
\vskip -0.5cm
\label{kien7}
\end{center}
\end{figure}

The intensity of the features determining the Fermi surface
should be calculated using the spectral function
$A({\bf k},\omega)$. Figure~\ref{kien3} only shows the eigenvalues of the mean-field
study, without incorporating the photoemission intensity of each state.
It is only when the strength of the coupling $U$ becomes very large  that the spectral weights for all the bands will be
equal. Otherwise, some of the bands will produce strong FS while others will produce only
weak magnetically-induced ``shadow'' features that are hard to observe, as already shown
in Fig.~\ref{kien4}. To better visualize the bands induced
by magnetic order, in Fig.~\ref{kien5} the mean-field
spectral function $A({\bf k},\omega)$ is presented along
high-symmetry directions in the BZ for the two-orbital model with the SK parameters.\cite{daghofer} For $U<U_{\rm c1}$, panel (a), the spectral weight
resembles the non-interacting band structure, i.e. there is negligible
spectral weight in the magnetically-induced bands. For $U_{\rm c1}<U<U_{\rm c2}$,
panel (b),
the bands become distorted and bands of magnetic origin develop particularly at the
locations in which a gap opens. There are other
bands still crossing the Fermi energy, thus
the system is metallic.
Finally for $U>U_{\rm c2}$,
panel (c), the gap is
complete and the magnetic bands are well developed, i.e., four peaks can
be observed in $A({\bf k},\omega)$ for almost all values of ${\bf k}$.
Figure~\ref{kien7}
shows the results for the LDA fitted hoppings,\cite{scalapino}
at $J$=$U/4$: here a similar qualitative discussion applies.
%
Other values of $J$ such as $J$=$0$ and $J=U/8$
(not shown) were also considered, and the results are qualitatively the same.

\subsection{Exact Diagonalization results}

To analyze the qualitative reliability of the mean-field results, we have performed
ED calculations on finite clusters. In our previous effort,\cite{daghofer}
it was discussed that due to the rapid growth of the
Hilbert space with the number of sites $N$, the largest cluster where
the two-orbital model can be exactly diagonalized has only $N$=$8$ sites. It is a tilted
$\sqrt{8}$$\times$$\sqrt{8}$ cluster, and when periodic
boundary conditions are implemented the available values
of the momenta are ${\bf k}=(0,0)$, $(\pm\pi/2,\pm\pi/2)$, $(0,\pi)$,
$(\pi,0)$, and $(\pi,\pi)$. This limited set of momenta
is not well suited to analyze band
dispersions in the BZ and the use of ``twisted'' boundary conditions  (see below)
for the 8-site cluster would require too high a computational effort.
However, via the study of spin correlations
it has been observed that the spin-striped magnetic order
characteristic of this model is also apparent in the even smaller $2\times 2$ cluster. For
this very small system, the limited number of available momenta can be enlarged by
implementing ``twisted'' boundary conditions (TBC), namely requesting
that $d(N_i+1)$=$e^{i\phi}d(1)$ where $N_i$ is the number of sites along the
$i$=$x$ or $y$ direction in the square cluster, and $\phi$ is an arbitrary phase.
With these TBC
the values of momenta allowed are now $k_i$=${2\pi n_i+\phi\over{N_i}}$ with $n_i$
ranging from 0 to $N_i-1$. Thus, we can calculate the spectral functions for
a variety of values of ${\bf k}$ using this TBC approach applied to the $2 \times 2$
cluster. While the very small size is still a serious limitation, note that
there are simply no other procedures
available to contrast the mean-field results against
exact results at intermediate couplings. Our goal using this limited size
cluster is merely to analyze if mean-field conclusions stand against exact results.


\begin{figure}[thbp]
\begin{center}
\includegraphics[width=8cm,height=8cm,clip,angle=0]{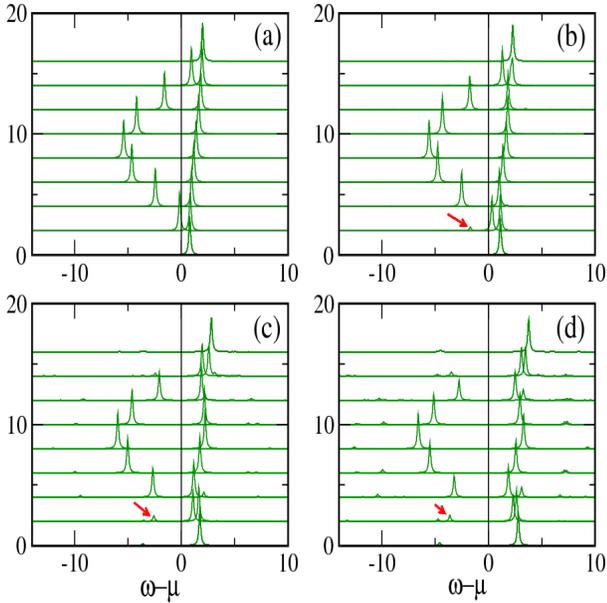}
\vskip -0.3cm
\caption{(Color online) Two-orbital model
spectral function along the $(0,0)$ to $(\pi,\pi)$ direction
in the extended FBZ for the LDA fitted
hoppings.\cite{scalapino}
(a) is for $U$=0.0, (b) for $U$=2.5, (c) for $U$=5.0, and (d) for $U$=8.0.
The Hund coupling is $J$=$U/4$. The method is ED and
the lattice is 2$\times$2 with TBC.
The arrows indicate the magnetic bands discussed in the text.}
\vskip -0.5cm
\label{doug_akju4}
\end{center}
\end{figure}

\begin{figure}[thbp]
\begin{center}
\vskip 0.5cm
\includegraphics[width=8cm,height=8cm,clip,angle=0]{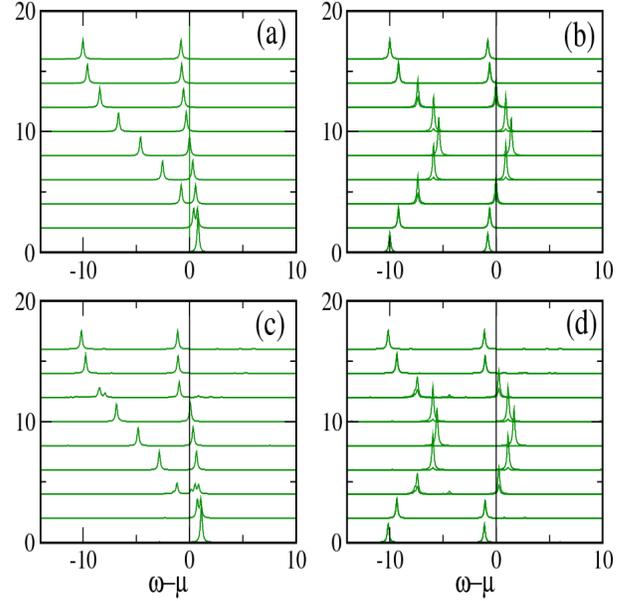}
\vskip -0.3cm
\caption{(Color online) Two-orbital model
spectral function along the $(0,0)$ to $(\pi,0)$ and $(\pi,0)$ to $(0,\pi)$
directions
in the extended FBZ for the LDA fitted
hoppings.\cite{scalapino}
(a,b) is for $U$=0, while (c,d) is for $U$=2.5.
The Hund coupling is $J$=$U/4$. The method is ED and
the lattice is 2$\times$2 with TBC. This figure shows that the results barely change between the two
values of $U$, suggesting the survival of a metallic state at nonzero $U$.}
\vskip -0.9cm
\label{doug_akju4_other}
\end{center}
\end{figure}

In Fig.~\ref{doug_akju4}, the spectral function
$A({\bf k},\omega)$ is presented along the main diagonal
of the extended BZ for different values of $U$ and with $J$=$U/4$.
These data have to be compared with the mean-field prediction shown in
Fig.~\ref{kien7}.
We present the results for $U$=$0$ for comparison and to demonstrate that the
correct dispersion is obtained in spite of the fact that the cluster is so small.
The finite values
of $U$ have been chosen to be in the magnetic-metallic region ($U$=$2.5$ and $5.0$) and
in the insulating region ($U$=$8$) according to the mean-field results. The main point
of this figure is to report the development of bands induced by magnetic order with increasing $U$. A
representative momentum for these magnetically-induced bands is highlighted
with an arrow in the figure.
With increasing $U$, the magnetic bands smoothly develop (concomitant with
the development of spin stripe order at short distances, as shown later) and at least
along the main-diagonal direction in the extended FBZ that should occur simultaneously
with the opening of a gap.
Thus, our first conclusion is that the extra weak features in the one-particle spectral
function predicted by mean-field due to
the magnetic order do appear in the ED results. At large $U$ there is
no doubt that a substantial gap is observed, as in the mean-field approach.
We also performed calculations with other values of $J$ such as
$J=0$ and $J=U/8$ (not shown), and the conclusions are qualitatively the same
as for $J$=$U/4$. Qualitatively similar conclusions were reached using
the SK hoppings.\cite{daghofer}

Let us analyze now other directions in momentum space. In Fig.~\ref{doug_akju4_other},
the $(0,0)$ to $(\pi,0)$ and $(\pi,0)$ to $(0,\pi)$ directions are investigated
at $U$=0 and 2.5. The results indicate negligible changes along these
directions by turning on $U$: the system appears to remain metallic.
However, as shown below, the NN spin-stripe order
in this small cluster is already robust at $U$=2.5.
Thus, these results are
compatible with the concept of a state simultaneously metallic and magnetically
ordered.
Moreover, by monitoring the
opening of the complete gap we found that the metal-insulator transition occurs
at a value of $U_{\rm c2}$ in good agreement with the mean-field predictions.
The existence of a
$U_{\rm c1}$ is a more complicated issue but it can be inferred from
the development of the magnetic bands in
$A({\bf k},\omega)$ which also occurs in a range of $U$ consistent with
mean-field.

\begin{figure}[thbp]
\begin{center}
\includegraphics[width=8cm,height=8cm,clip,angle=0]{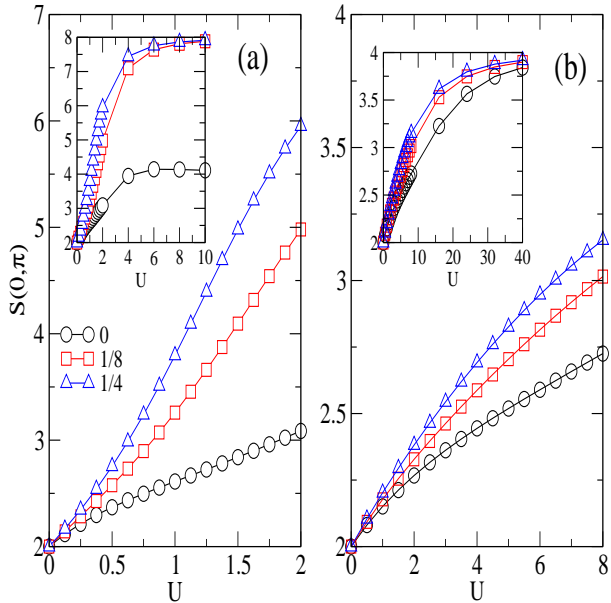}
\vskip -0.3cm
\caption{(Color online) The magnetic structure factor at ${\bf k}$=$(0,\pi)$
vs. $U$ calculated by ED of an 8-site cluster
 for the values of $J$ indicated. (a) is for the SK hoppings;\cite{daghofer}
and (b) for the LDA fitted
hoppings.\cite{scalapino}
The insets show the results in a more extended range of $U$.}
\vskip -0.3cm
\label{struc_s8}
\end{center}
\end{figure}

In Fig.~\ref{struc_s8} we present the magnetic structure factor $S$ for
${\bf k}$=$(0,\pi)$ which is the value of the momentum for which it has a
maximum (degenerate with $(\pi,0)$ for this small cluster);
panel (a) shows results for the SK hoppings\cite{daghofer}
while panel (b) is for the LDA fitted hoppings.\cite{scalapino}
To reduce finite-size effects results for the
$N=8$ cluster are presented, although the results in the $2\times 2$ cluster are qualitatively
similar. For large $U$, the monotonic increase of $S$ with $U$ agrees with
the mean-field results. At small and intermediate $U$, the spin stripe correlations
are robust and for these small clusters this is equivalent to long-range order.
But the apparent lack of a gap at $U$=2.5 along particular directions in momentum
space (discussed before) lead us to believe that the ED results are compatible
with a metallic and magnetic phase at intermediate $U$.



\subsection{VCA results}

\begin{figure}[thbp]
\begin{center}
\includegraphics[width=8cm,clip,angle=0]{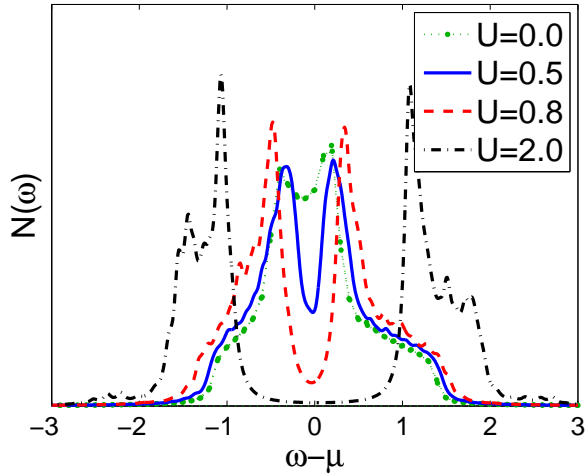}
\vskip -0.3cm
\caption{(Color online)  VCA calculated density of states for different values of $U$ in the
metallic (good metal and pseudogap) and insulating regimes with the SK hopping parameters using
$pd\pi/pd\sigma$=$-0.2$ and $J$=$U/4$.
}
\vskip -0.3cm
\label{dos}
\end{center}
\end{figure}

In this final subsection, numerical results for the spectral functions
and the density of states obtained with the Variational Cluster Approximation (VCA)
technique\cite{Aic03,Pot03} are presented.
This method embeds the ED solution of a small $2\times 2$
cluster into a very large system of a size comparable to the
$100\times 100$ momentum points used in the mean field, and thus
interpolates between the results obtained independently by the ED and
mean-field approaches.
The VCA results discussed here are for the SK parameters.\cite{daghofer}
Figure~\ref{dos} shows the VCA density of states. The
behavior with increasing $U$ is remarkably similar to that obtained in
our mean-field calculations presented in Fig.~\ref{kien2}~(b). Metallic, pseudogap,
and insulating regimes can be clearly observed.

\begin{figure}[thbp]
\begin{center}
\includegraphics[width=6.5cm,clip,angle=0]{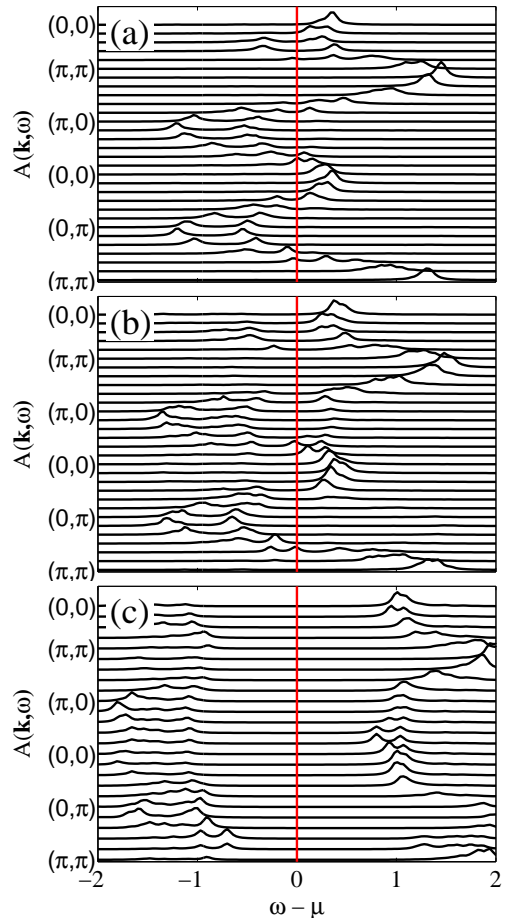}
\vskip -0.3cm
\caption{(Color online)  VCA calculated spectral functions along high symmetry directions in
the BZ for the SK hopping parameters with
$pd\pi/pd\sigma$=$-0.2$, $J$=$U/4$, and wavevector $(\pi,0)$.
(a) $U$=0.5;
(b) $U$=0.8; (c) $U$=2.0.
}
\vskip -0.9cm
\label{vca}
\end{center}
\end{figure}

The results for $A({\bf k},\omega)$ calculated with VCA are shown in
Fig.~\ref{vca}. Once again, a remarkable quantitative agreement with
the mean-field results of Fig.~\ref{kien5} is found.

\section{Conclusions}

In this investigation, the mean-field technique was applied to multi-orbital
Hubbard models for the Fe-pnictides. Varying $U$, three regions were observed. At
small coupling, the results are as in the non-interacting limit. In the other extreme of
very large $U$, the ground state has a robust gap and the magnetic spin-stripe
order parameter is large. The main result of our effort is the presence
of an intermediate $U$ coupling regime where magnetic spin-stripe order is shown to be
compatible with a metallic ground state due to band overlaps. This state has
similar characteristics as the parent compounds of the Fe-pnictide
superconductors.
Although further theoretical work is still needed to
firmly establish the existence of this interesting intermediate coupling regime,
for the case of two orbitals our conclusions were tested using
the ED and VCA methods, and the results are compatible with
mean-field.

The analysis of the intermediate $U$ regime allowed us to predict the results for
angle-resolved  photoemission experiments. Interesting anisotropies manifested
as arcs at the Fermi surface. New bands of magnetic origin were also discussed.
The two- and four-orbital models lead to similar results
in this context.
Future work will address the optical properties
of the intermediate coupling regime, and the superconducting state
that may arise from its doping.

\section{Acknowledgments}
This work was mainly supported by the NSF grant DMR-0706020 and the
Division of Materials Science and Engineering, U.S. DOE, under contract
with UT-Battelle, LLC.
Computation for part of the work described in this paper was supported by the
University of Southern California Center for High Performance Computing
and Communications. S.H. and K.T. acknowledge financial support by the National
Science Foundation under grant DMR-0804914 and the Department of Energy
under grant DE-FG02-05ER46240.


\begin{thebibliography}{10}
\bibitem{noteorder}
By symmetry consideration, the state ordered at $(\pi,0)$ will give exactly the
  same result, with interchange of $xz$ and $yz$ orbitals.
\end{thebibliography}

\end{document}